\documentclass[aps,prd,twocolumn,showpacs,showkeys,amsmath,amssymb]{revtex4}
\usepackage{graphicx}% Include figure files
\usepackage{dcolumn}% Align table columns on decimal point
\usepackage{bm}% bold math
\usepackage{mathtext}
\usepackage[dvips]{epsfig}
\usepackage{epsfig}
\usepackage{graphics}
\usepackage{subfigure}
\usepackage{mathrsfs}
\usepackage{amssymb,latexsym}
\newcommand{\seq}{\begin{subequations}}
\newcommand{\sen}{\end{subequations}}
\newcommand{\eq}{\begin{eqnarray}}
\newcommand{\en}{\end{eqnarray}}
\def\shiftleft#1{#1\llap{#1\hskip 0.04em}}
\def\shiftdown#1{#1\llap{\lower.04ex\hbox{#1}}}
\def\thick#1{\shiftdown{\shiftleft{#1}}}
\def\b#1{\thick{\hbox{$#1$}}}
\begin{document}
\title{Quasielastic $\b{\rho^0}$ electroproduction on the proton at 
intermediate energy:\\ Role of scalar and pseudoscalar meson exchange} 

 \author{I.T. Obukhovsky$^1$,
Amand Faessler$^2$,
D.K. Fedorov$^1$,
Thomas Gutsche$^2$, 
Valery E. Lyubovitskij$^2$
\footnote{On leave of absence
from Department of Physics, Tomsk State University,
634050 Tomsk, Russia}, 
V.G. Neudatchin$^1$, 
L.L. Sviridova$^1$
\vspace*{1.2\baselineskip}}

\affiliation{$^1$Institute of Nuclear Physics, Moscow
State University,119991 Moscow, Russia
\vspace*{1.2\baselineskip} \\
\hspace*{-1cm} 
$^2$Institut f\"ur Theoretische Physik,
Universit\"at T\"ubingen,\\
Kepler Center for Astro and Particle Physics, \\
Auf der Morgenstelle 14, D--72076 T\"ubingen, Germany
\vspace*{0.3\baselineskip}\\}

%\date{\today}

\begin{abstract}
  
\noindent
We present a detailed analysis of $\rho^0$ electroproduction at 
intermediate energy for quasi-elastic knockout kinematics. 
The approach is based on an effective Lagrangian which generates exchanges of 
scalar, pseudoscalar, axial-vector and tensor mesons. The specific role of 
mesons with different values for $J^{PC}$ is analyzed. We show that the 
$\pi^0$ exchange amplitude and its interference terms with $\eta(\eta^\prime)$ 
and $f_1$ exchanges dominate in the transverse part of the cross section 
$\sigma_{\scriptscriptstyle T}$. The main role plays the $\pi^0\to\rho^0$ M1 
spin transition when coupling the virtual photons to the nucleon. In contrast, 
the longitudinal cross section $\sigma_{\scriptscriptstyle L}$ is generated by 
a series of scalar meson exchanges. To extract the dominant term more detailed 
information on the inner structure of scalar mesons is required. It turns out 
that recent data of the CLAS Collaboration on $\sigma_{\scriptscriptstyle L}$ 
and $\sigma_{\scriptscriptstyle T}$ can be described with reasonable accuracy 
if one proposes the quarkonium structure for the heavy scalar mesons 
($M\gtrsim$ 1.3 GeV). On this basis the differential cross sections 
$d\sigma_{\scriptscriptstyle L}/dt$ and $d\sigma_{\scriptscriptstyle T}/dt$ 
are calculated and compared with the latest CLAS data.

\end{abstract}



\maketitle

%%%%%%%%%%%%%%%%%%%%%%%%%%%%%%%%%%%%%%%%%%%%%%%%%%%%%%%%%%%%%%%%%%%%%%
\section{Introduction.}
%%%%%%%%%%%%%%%%%%%%%%%%%%%%%%%%%%%%%%%%%%%%%%%%%%%%%%%%%%%%%%%%%%%%%%%%

Recent data of the CLAS Collaboration~\cite{clashad,clasmor} on 
electroproduction of the $\rho^0$ meson with separation of longitudinal 
$\sigma_{\scriptscriptstyle L}$ and transverse $\sigma_{\scriptscriptstyle T}$ 
parts of the cross section open up new possibilities in the study of the 
production mechanism. Experimental results on $\sigma_{\scriptscriptstyle L}$ 
and $\sigma_{\scriptscriptstyle T}$ obtained at relatively large photon 
virtualities 
of 1.5 $\lesssim Q^2\lesssim$ 4 GeV$^2/c^2$ and at invariant energies $W\sim$ 
2 -- 3 GeV (i.e. above the resonance region) supplement the known data 
of the CLAS Collaboration on $\rho^0$ photoproduction~\cite{clasbat} since 
they contain new information on the $Q^2$--dependence of the cross sections. 
At large $Q^2$ some contributions to the cross section are small and in  
first approximation they can be excluded from the consideration. For example, 
at $Q^2\gtrsim$ 1.5 GeV$^2$/c$^2$ contributions of intermediate baryon states 
($N,\,N^*$, etc.) to the cross section are sufficiently suppressed in 
comparison to meson exchange contributions due to form factors encoding finite 
size effects in the $\gamma NN$, $\gamma NN^*$, $\rho NN$ and $\rho NN^*$ 
vertices. 

An important feature of $\rho^0$ electroproduction is that different meson 
exchange mechanisms dominate in the longitudinal and transverse cross 
sections. This phenomenon allows to consider the contributions of these mesons 
separately, while the data on only the full cross section 
$\sigma=\sigma_{\scriptscriptstyle T}+\epsilon\,\sigma_{\scriptscriptstyle L}$ 
do not permit such a possibility. For example, data of the JLAB $F_\pi$ 
Collaboration on electroproduction of charged pions~\cite{huber,horn,volmer} 
allowed to deduce the charged pion form factor using the dominance of the pion
$t$-pole contribution to the longitudinal part of the differential cross 
section  $d\sigma_{\scriptscriptstyle L}/dt$. Note, that the transverse part 
$d\sigma_{\scriptscriptstyle T}/dt$ is dominated by the $\rho$--meson pole, 
and the JLab 
data~\cite{huber,horn,volmer} on $d\sigma_{\scriptscriptstyle T}/dt$ 
allow to perform an independent study of the $\rho$--meson exchange 
amplitude~\cite{faess}. It would be impossible to study both phenomena on 
the basis of the total cross section only.

These new possibilities for a detailed experimental study of the reaction
$\gamma^*_{\scriptscriptstyle L,T}+p\to\rho^0+p$ in quasi-elastic kinematics
can be used to gain insight into the structure of the meson cloud of the 
nucleon and into the quark origin of the electromagnetic properties of 
neutral mesons. Our present work is devoted to the theoretical study of this 
reaction in the context of these new possibilities that distinguish recent 
electroproduction experiments~\cite{clashad,clasmor} from older 
ones~\cite{cassel}.

In the case of the $\gamma^*_{\scriptscriptstyle L,T}+p\to\rho^0+p$ reaction
until recently there were only integral data (integral characteristics over
the $t$ variable -- the squared momentum transfer to the proton -- see e.g.
Ref.~\cite{clashad})
which we have used as a basis for our consideration.
Very recently data on the differential cross sections were also published
by the CLAS Collaboration~\cite{clasmor} which allows for a further detailed test
of the theory presented here.

The differential cross sections 
$d\sigma_{\scriptscriptstyle T}/dt$ and $d\sigma_{\scriptscriptstyle L}/dt$ 
are more informative: for small values of $t$ near the kinematical 
threshold $t-t_{min}\sim$ 0 (i.e. in the region of quasi-elastic meson 
knockout) the nearby $t$--channel pole dominates in each of these cross 
sections, which generates the forward peak of electroproduction in the 
absorption of either a longitudinal ($\gamma^*_{\scriptscriptstyle L}$) or 
transverse ($\gamma^*_{\scriptscriptstyle T}$) virtual photon. In the case of 
$\gamma^*_{\scriptscriptstyle L,T}+p\to\rho^0+p$ the pion $t$--pole contribution 
dominates in $\sigma_{\scriptscriptstyle T}$ and $t$ poles of the lightest
scalar mesons dominate in $\sigma_{\scriptscriptstyle L}$. 
The dominant $t$ pole gives also a main contribution to the corresponding 
integrated cross section ($\sigma_{\scriptscriptstyle T}$ or 
$\sigma_{\scriptscriptstyle L}$), but contributions of heavy-meson exchanges 
(mesons with masses close to and above 1 GeV) are also important. 
Sometimes these corrections are quite significant since they interfere 
with the leading contributions of the light mesons. 

Nevertheless, as the first step, the data on $\sigma_{\scriptscriptstyle L}$ and 
$\sigma_{\scriptscriptstyle T}$~\cite{clashad} for the reaction 
$\gamma^*_{\scriptscriptstyle L,T}+p\to\rho^0+p$ can already be used to give an 
estimate for the dominant contributions. 
The additional (in comparison to electroproduction of charged mesons) 
selection rule connected with charge parity (C) conservation in the
$\gamma\to\rho^0$ transition allows this procedure. Latter conservation law
reduces the number of exchange diagrams to be considered. Here we also can 
use the models of vector meson dominance (VMD) and tensor meson dominance 
(TMD)~\cite{renner} for an estimate of the vertex constants in 
the $t$--channel pole terms~\cite{oh} using the underlying idea of 
charge universality. 

In the reaction $\gamma^*_{\scriptscriptstyle L,T}+p\to\rho^0+p$ both the 
virtual photon and $\rho^0$ have the same (negative) charge parity, i.e. 
$J^{PC}=1^{--}$. Such a reaction cannot be considered as a true quasi-elastic 
knockout process, since here $\rho^0$-- meson exchange is forbidden due to 
$C$--parity conservation. Therefore, such a ``knockout'' can proceed  
due to the conversion of a meson from the nucleon meson cloud into the final 
$\rho^0$ meson or due to the $\gamma \to \rho^0$ transition in a diffraction 
process. Since pion exchange ($J^{PC}=0^{-+}$) is allowed here the pion pole 
must dominate in the region of small $|t|$ (i.e. at small angles in 
quasi-elastic kinematics). However, the pion contribution only dominates in the
transverse part of the cross section $\sigma_{\scriptscriptstyle T}$ due to 
the $M1$ spin transition $\pi^0\to\rho^0$. At the same time, the pion pole 
contribution to $\sigma_{\scriptscriptstyle L}$ is negligibly small even at 
values of $t\sim t_{min}$ -- opposite to the situations of charged meson
knockout processes. Pomeron exchange ($C=+$) is allowed but its 
contribution to $\sigma_{\scriptscriptstyle L}$ is too small in the 
region of invariant energies $W\sim$ 2 -- 3 GeV considered when compared to
the summation of the $t$-pole terms of other low-mass mesons. 
 
For invariant energies $W$ slightly above the baryon resonance region and for
high virtualities of the photon ($Q^2 \sim$ 1.5 -- 4 GeV$^2$/c$^2$ in the JLAB 
experiments) it is sufficient to take into account the exchanges of neutral 
pseudoscalar mesons and additionally from the three nonets $J^{PC}=J^{++}$ 
with $J=$ 0,1,2 corresponding to the first orbital 1P excitation of the vector 
nonet (see Table I). Note, that we only consider mesons with positive charge 
parity (C=+) which give a contribution to the quasi--elastic knockout of 
$\rho^0$ supplemented either by spin flip ($M1$--transitions 
$^1S_0$ $\to$ $^3S_1$ without changing the spatial $P$-parity) or 
deexcitation of the orbital $1P$ 
state ($E1$--transitions $^3P_J$ $\to$ $^3S_1$ with change of the $P$-parity). 
In the first approximation one can neglect the highly excited meson nonets, 
because the corresponding orbital matrix elements 
of the transitions $2S\to 1S$, $2P\to 1S$, etc. must be suppressed 
in comparison to the $1S\to 1S$ and $1P\to 1S$ contributions.  

Starting with energies of $W\sim$ 5 -- 10 GeV and above the electroproduction 
cross section is suitably described
in the framework of Regge phenomenology, which gives a reasonable description 
in a wider region of the variable $t$ than the $t$-pole approximation, -- 
up to the region of hard collisions where partonic degrees of freedom become 
manifest. Then the most convenient description of the hadronic 
processes can be done in terms of generalized partonic distributions (see, 
e.g.~\cite{clashad,clasmor,guidal,kaskul}). 

In a series of works~\cite{laget1,laget2,laget3,cano} the Regge phenomenology 
has been extended to the description of meson electro-- and photoproduction 
cross sections at lower energies of about $W\sim$ 2 -- 3 GeV. In this 
approach meson propagators in the exchanged diagrams are substituted by 
amplitudes of the corresponding Regge trajectories (R). The Pomeron trajectory 
(P) contribution is also included in the total sum. Here the coupling constants
and form factors of low--energy hadron physics are used for the $\gamma R\rho$ 
and  $RNN$ vertices. For the $\gamma R\rho$ vertex the dominance of vector 
mesons is used and a corresponding form factor is calculated in terms 
of the $q\bar q$ loop in the Landshoff-Donnachie approach~\cite{landshoff}. 
Since the vertex coupling constants (excluding the $\pi NN$ coupling) 
are only known with a low precision an additional free 
parameter~\cite{laget3,cano} is introduced into the amplitude which is 
normalized by data on $\rho^0$ photoproduction. 

We should stress that in the energy region of $W\sim$ 2 -- 3 GeV an equally 
good description of meson photo-- and electroproduction cross sections
can also be obtained on the basis of the usual pole approximation -- also 
using phenomenological vertex form factors~\cite{oh,faess,obukh,neud2}. 
An advantage of the pole approximation is that the starting point is set up
by effective Lagrangians. Therefore the momentum--spin structure of the meson 
vertices can be consistently taken into account which also defines the energy 
dependence of cross sections in the case of higher meson spins. 

In Ref.~\cite{clashad} a reasonable description of the CLAS data on  $\rho^0$ 
electroproduction has been obtained in the framework of a Regge 
model~\cite{laget3,cano} using the dominance of $\pi$, $\sigma$ 
and $f_2$ trajectories. The tensor $f_2$ meson was implemented as an isoscalar 
meson with positive $C$--parity and $J=$ 1 in accordance with the hypothesis 
that pomeron and $f_2$ trajectories are proportional. Also, an additional 
multiplier $\kappa_{f_2}$ was introduced to rescale the contribution of 
the $f_2$ trajectory relative to the one of the pomeron. The size of 
$\kappa_{f_2}$ was normalized to experimental data on $\rho^0$ 
photoproduction~\cite{clasbat}. As it turns out a significant enhancement of 
the $f_2$ trajectory contribution as compared to the pomeron trajectory 
($\kappa_{f_2}=$ 9) is required. The $\sigma$ meson exchange has also been 
enhanced, because in~\cite{laget3} a large value for the $\rho\sigma\gamma$ 
coupling was used ($g_{\rho\sigma\gamma}\approx$ 1, see details 
in~\cite{cano}). Data on the 
$\rho\to\pi^0\pi^0\gamma$ decay width~\cite{pdg,snd} can be explained 
using a much smaller value for the $g_{\rho\sigma\gamma}$ coupling~\cite{snd}. 

All this has been analyzed in Ref.~\cite{oh} where the description of data 
on $\rho$ photoproduction obtained earlier by~\cite{laget3} has been 
reconsidered. An alternative approach was proposed, where the amplitude of
$\rho^0$ photoproduction was represented by the sum of the $t$--pole  
contributions from {\it physical} meson exchanges with coupling constants 
normalized to independent data ($s$- and $u$-pole contributions were taken 
into account as well). A good description of photoproduction data has been 
achieved in both approaches~\cite{laget3} and \cite{oh}. It seems that the
$\sigma$ and $f_2$ mesons showing up in the Regge model~\cite{laget3} are only 
{\it effective} degrees of freedom giving a useful parametrization of the 
total contribution by exchange of {\it physical} mesons listed in  Table I. 

In the present paper we also pursue an alternative description of the
data on $\rho^0$ electroproduction similar to the approach of Ref.~\cite{oh}.
We start with phenomenological Lagrangians to calculate the cross sections 
$\sigma_{\scriptscriptstyle L}$ and $\sigma_{\scriptscriptstyle T}$ in the 
$t$--pole approximation for the set of mesons displayed in Table I (see also 
Fig.1). In our calculation we use the coupling constants and form factors 
supported by and deduced from data and which mostly coincide 
with those already used in Ref.~\cite{oh} in the description 
of photoproduction data. The comparison of the theoretical results with latest 
data of the CLAS Collaboration~\cite{clashad} allows to determine the set 
of dominant meson exchanges corresponding to {\it physical} particles.  

It will be found that in the description of the transverse cross section 
$\sigma_{\scriptscriptstyle T}$ the pion contribution and the summed 
contribution of pseudovector mesons ($f_1,\,f_1^\prime,\,a_1$) and other 
pseudoscalars ($\eta,\,\eta^\prime$) in interference with the pion piece
play the major role. In addition, the summed contribution of mesons with 
positive $P$--parity ($f_0,\,f_2$, $a_0,\,a_2$) is practically not visible in 
$\sigma_{\scriptscriptstyle T}$ above such a background. In contrary, 
mesons with positive $P$--parity give a significant contribution to 
$\sigma_{\scriptscriptstyle L}$. However, if we use for all scalar $f_0$
mesons of Table I electromagnetic coupling constants $g_{\rho f_0\gamma}$ 
normalized to the known radiative decay widths of the $\sigma$ and $f_0(980)$ 
mesons, the contribution of all $f_0$,$a_0$,$f_2$ and $a_2$ mesons will be 
not sufficient to explain the data on $\sigma_{\scriptscriptstyle L}$. 

In the final part of the paper we discuss different scenarios for overcoming
these difficulties. It is shown that the problem can be solved 
using the set of known scalar mesons (i.e. without inclusion of 
possible exotic states) if the set of heavy scalars 
($f_0(1370)$, $f_0(1500)$ and $f_0(1710)$ or at least two states from 
this group) are considered as $^3P_0$ states in the $\bar qq$ spectrum. 
According to the quark model such states should have relatively large 
radiative decay widths $\Gamma_{f_0\to\gamma\rho}\gtrsim$ 
100 KeV~\cite{kalash1} resulting in large coupling constants 
$\rho f_0\gamma$ with 
$g_{\rho f_0\gamma}>>g_{\rho\sigma\gamma}$ ($\bar qqg$ hybrid models 
with anomalous large widths~\cite{kalash2} and models~\cite{soy,kissl} with
large values for the $\rho\sigma\gamma$ or $\sigma NN$ coupling
also do not contradict data, but one can proceed without these assumptions). 
Furthermore, in addition to the included $\sigma$ meson exchange 
a non-correlated $2\pi$ exchange must also contribute to the 
cross section (recall that in Ref.~\cite{oh} it was shown that
this mechanism plays an appreciable role in the $\rho^0$ photoproduction at 
low energy). 

We present our final results on the integrated $\sigma_{\scriptscriptstyle L/T}$ 
and differential $d\sigma_{\scriptscriptstyle L/T}/dt$ cross sections where the
previously indicated corrections for the coupling constants 
$g_{\rho f_0\gamma}$ of two heavy mesons $f_0(1370)$ and $f_0(1500)$ are
included.
The results for $\sigma_{\scriptscriptstyle L/T}$ are compared to
the latest data of the CLAS Collaboration~\cite{clashad,clasmor}. Theoretical
curves are in agreement with data within experimental errors.
The results for $d\sigma_{\scriptscriptstyle L/T}/dt$ appear to be in a good
agreement with the new CLAS data~\cite{clasmor} as illustrated for
several experimental bins in the range of 1.9 $<\,Q^2\,<$ 2.2 GeV$^2/c^2$.
A full analysis of all the experimental bins (more than 50 kinematical regions
from $Q^2=$ 1.6 GeV$^2/c^2$ and $x_B=$ 0.16 to $Q^2=$ 5.6 GeV$^2/c^2$ and
$x_B=$ 0.7) will be presented in a further paper in its own right.

%%%%%%%%%%%%%%%%%%%%%%%%%%%%%%%%%%%%%%%%%%%%%%%%%%%%%%%%%%%%%%%%%%%%%%
\section{Framework}
%%%%%%%%%%%%%%%%%%%%%%%%%%%%%%%%%%%%%%%%%%%%%%%%%%%%%%%%%%%%%%%%%%%%%%

%%%%%%%%%%%%%%%%%%%%%%%%%%%%%%%%%%%%%%%%%%%%%%%%%%%%%%%%%%%%%%%%%%%%%%
\subsection{$t$--pole contributions due to exchange of mesons 
with positive $C$--parity} 
%%%%%%%%%%%%%%%%%%%%%%%%%%%%%%%%%%%%%%%%%%%%%%%%%%%%%%%%%%%%%%%%%%%%%%%%

\subsubsection{Scalar meson exchange ($S=f_0$, $a_0$)}

We start with the effective Lagrangian 
\begin{eqnarray}
{\cal L}_{\rho {\scriptscriptstyle S}\gamma}(x)=
\frac{eg_{\rho S\gamma}}{4M_\rho}S(x)F_{\mu\nu}(x)\rho^{\mu\nu}(x),\nonumber\\
F_{\mu\nu}=\partial_\mu A_\nu-\partial_\nu A_\mu,\quad
\rho_{\mu\nu}=\partial_\mu \rho_\nu-\partial_\mu \rho_\nu,
\label{1}
\end{eqnarray}
\begin{equation}
{\cal L}_{{\scriptscriptstyle SNN}}(x)=g_{{\scriptscriptstyle SNN}}
S(x)\overline{N}(x)N(x)
\label{2}
\end{equation}
which generates matrix elements due to the exchange of scalar mesons  
$S=f_0,\,a_0$. The respective invariant amplitude is
\begin{multline}
T_{\scriptscriptstyle S}(s,s^\prime,\lambda,\lambda_\rho)=
e\,\frac{g_{\rho {\scriptscriptstyle S}\gamma}}
{M_\rho}g_{{\scriptscriptstyle SNN}}
\frac{g_{\mu\nu}k^\prime\cdot q-k^\prime_\nu q_\mu}{k^2-M_{f_0}^2+i0}\\
\times{\epsilon_{\scriptscriptstyle V}^{(\lambda_\rho)\mu}}^*\!(k^\prime)\,
\epsilon^{(\lambda)\nu}\!(q)\overline{u}(p^\prime,s^\prime)u(p,s),
\label{3}
\end{multline} 
where $\epsilon^{(\lambda)}_\nu(q)$ and
$\epsilon_{{\scriptscriptstyle V}\,\mu}^{(\lambda_\rho)}(k^\prime)$ 
are the polarization vectors of photon and $\rho^0$ meson, respectively. 
They satisfy completeness relations in the subspace orthogonal to the 
4--momentum: 
\begin{eqnarray}
\sum_{\lambda\!=\!0,\pm1}(\!-\!1)^\lambda\epsilon^{(\lambda)}_\mu(q)
{\epsilon^{(\lambda)}_\nu}^*(q)&=&
g_{\mu\nu}-\frac{q_\mu q_\nu}{q^2},\nonumber\\
\sum_{\lambda_\rho\!=\!0,\pm1}
\epsilon_{{\scriptscriptstyle V}\,\mu}^{(\lambda_\rho)}\!(k^\prime)
{\epsilon_{{\scriptscriptstyle V}\,\nu}^{(\lambda_\rho)}}^*\!(k^\prime)
&=&-\left[g_{\mu\nu}-\frac{k^\prime_\mu k^\prime_\nu}{M_\rho^2}\right].
\label{4}
\end{eqnarray} 
Further, the invariant amplitude~(\ref{3}) is modified as usually by 
introducing vertex form factors
\begin{equation}
g_{\rho {\scriptscriptstyle S}\gamma}\to g_{\rho {\scriptscriptstyle S}\gamma}
{\cal F}_{\rho {\scriptscriptstyle S}\gamma}(Q^2,t),\quad 
g_{S{\scriptscriptstyle NN}}\to
g_{S{\scriptscriptstyle NN}}{\cal F}_{S{\scriptscriptstyle NN}}(t),
\label{5} 
\end{equation} 
where 
\eq 
{\cal F}_{\rho {\scriptscriptstyle S}\gamma}(Q^2,t) \equiv 
{\cal F}_1(Q^2){\cal F}_2(t)\,, \hspace*{.25cm}  
{\cal F}_{S{\scriptscriptstyle NN}}(t) \equiv {\cal F}_3(t)\,, 
\en  
$t=k^2$ and $Q^2=-q^2$. 
The substitution~(\ref{5}) is equivalent to  
a nonlocal form of the interaction vertices~\cite{efim,ivanov,fae}, i.e. to
the following modification of the local Lagrangians~(\ref{1}) and (\ref{2}): 
\begin{eqnarray}
{\cal L}_{\rho {\scriptscriptstyle S}\gamma}(x)&\to&
{\cal L}^{NL}_{\rho {\scriptscriptstyle S}\gamma}(x)=
\frac{eg_{\rho{\scriptscriptstyle S}\gamma}}{4M_\rho} 
\int \! d^4y \int \! d^4z \Phi_1(z^2) \Phi_2(y^2) \nonumber\\
&\times&
S(x+y)F_{\mu\nu}(x+z)\rho^{\mu\nu}(x)\,,
\nonumber\\
{\cal L}_{{\scriptscriptstyle SNN}}(x)&\to&
{\cal L}^{NL}_{{\scriptscriptstyle SNN}}(x)
=g_{{\scriptscriptstyle SNN}}\int \! d^4y \, \Phi_3(y^2)
\nonumber\\
&\times&S(x+y)\overline{N}(x)N(x),
\label{6}
\end{eqnarray}
where $\Phi_i(y^2)$ are relativistic invariant vertex functions. In momentum 
space ${\cal F}_i(k^2)=\int \Phi_i(y^2)e^{ik\cdot y}d^4y$ defines the 
corresponding vertex form factor.   

The factors $g_{\rho f_0\gamma}$ and
${\cal F}_{\rho f_0\gamma}(Q^2,t)$ in the effective 
Lagrangian~(\ref{1}) should correspond to a specific physical process 
in the vertex. In particular, they should take into account the
VMD transition $\gamma\to\rho^0$ with further diffractive scattering 
of the $\rho^0$ --- in accordance with the diagram shown in Fig.~2a. 
An analogous process should contribute to the vertex $\rho\pi\gamma$ 
(see Fig.~2b) with the difference that here the spin $M1$ transition 
$\pi^0\to\rho^0$ is understood. As a result the dependence of the form factors 
${\cal F}_{\rho f_0\gamma}(Q^2,t)$ and ${\cal F}_{\rho\pi\gamma}(Q^2,t)$ 
on the virtuality $Q^2$ of the photon is described by the propagator of the 
vector meson $\frac{1}{Q^2+M_V^2}$ in the first approximation. The dependence 
on $t$ involves the specific scale $\Lambda^{-1}$ corresponding to the size of 
the interaction volumes in the transitions $\rho^0+f_0\to\rho^0$ and 
$\omega+\pi\to\rho^0$ (see below). 

The magnitudes of the $\rho S\gamma$ coupling constants can be estimated 
using the radiative decay width of the scalar meson $f_0$ with
\begin{equation}
\Gamma_{f_0\!\to\!\gamma\rho}=
\alpha\,\frac{g_{\rho f_0\gamma}^2}{M_{\rho}^2}
\left(\frac{M_{f_0}^2-M_\rho^2}{2M_{f_0}}\right)^3,
\label{7}
\end{equation} 
while for the lightest scalar meson $f_0(600)\equiv\sigma$ the decay width 
for $\rho^0\to\gamma+\sigma$ 
\begin{equation}
\Gamma_{\rho\!\to\!\gamma\sigma}=
\frac{\alpha}{3}\frac{g_{\rho\sigma\gamma}^2}{M_{\rho}^2}
\left(\frac{M_{\rho}^2-M_{\sigma}^2}{2M_{\rho}}\right)^3
\label{8}
\end{equation} 
is used. In Eqs.~(\ref{7}) and (\ref{8}) we use coupling constants 
$g_{\rho f_0\gamma}$ fixed on the mass--shell ($Q^2=$ 0, $t=M_{f_0}^2$). 
Therefore, the form factor (\ref{5}) for the $\rho f_0\gamma$ vertex 
must be normalized at $Q^2=0$ as 
\begin{equation}
{\cal F}_{\rho {\scriptscriptstyle S}\gamma}(Q^2=0,t=M_{f_0}^2)=1 \,. 
\label{5a}
\end{equation}  
The $\sigma NN$ form factor is normalized according to 
\begin{equation}
{\cal F}_{\sigma {\scriptscriptstyle NN}}(t=0)=1,
\label{5b}
\end{equation}
since the constant $g_{\sigma{\scriptscriptstyle NN}}$ is defined 
from $NN$ scattering data at low energies in the limit $t\to$ 0.

According to the data of the SND Collaboration~\cite{snd} the width of  
$\rho^0$ decay into the channel with the lightest scalar meson $\sigma$ is 
sufficiently large: $\Gamma_{\rho\!\to\!\gamma\sigma}\approx$ 2.83 keV. 
Estimates of the decay width $\Gamma_{f_0\!\to\!\gamma\rho}$ are also known 
for heavier scalar mesons, e.g. for $f_0(980)$ ($\sim$ 3.4 keV) obtained 
in the framework of the molecular $K\bar K$ model 
$f_0(980)$ \cite{lyub,kalash1}. In both cases we obtain quite similar 
predictions for the coupling constants: 
\begin{eqnarray}
\quad g_{\rho\sigma\gamma}/M_{\rho}= 0.25/M_{\rho}=0.32\,GeV^{-1},\nonumber\\
\quad g_{\rho f_0\gamma}/M_{\rho}\approx 0.21/M_{\rho}=0.27\,GeV^{-1}.
\label{9}
\end{eqnarray} 
For the present purposes we use a unique value $g_{\rho\sigma\gamma}=$ 0.25 
for all scalar mesons in the calculation of the total exchange contribution 
involving $f_0(600)$, $f_0(980)$, $f_0(1370),$ $f_0(1500)$ and $f_0(1710)$. 

Presently no data are available to constrain the $f_0NN$ couplings except for 
the lightest scalar $\sigma=f_0(600)$. Here we take the value 
$g^2_{\sigma NN}/4\pi\simeq$ 4 $\div$ 8 commonly used in boson-exchange models 
of the $NN$ interaction. As a rough estimate for
of the contribution of the higher mass $f_0$ exchanges we use the common value 
$g_{f_0 NN}=$$g_{\sigma NN}=$ 10. We note that $\sigma_T$ is only weakly 
sensitive to even significant variation of the constants 
$g_{f_0 NN}$. Only $\sigma_L$ allows to search for an averaged contribution 
of $f_0$ meson exchanges. 

\bigskip

\subsubsection{Pseudoscalar meson exchange 
($S_5=\pi^0$, $\eta$, $\eta^\prime$)}

The $t$--pole contribution due to pseudoscalar meson exchange 
is described as 
\begin{multline} 
T_{\pi^0}(s,s^\prime,\lambda,\lambda_\rho)=
-\frac{e g_{\rho\pi\gamma}}{2m_{\scriptscriptstyle N} M_{\rho}}
{\cal F}_{\rho\pi\gamma}(Q^2,t) 
g_{\pi{\scriptscriptstyle NN}}{\cal F}_{\pi{\scriptscriptstyle NN}}(t)\\
\times\varepsilon^{\mu\nu\alpha\beta}\,
\frac{\epsilon^{(\lambda)}_\mu\!(q)q_\nu\,
{\epsilon_{{\scriptscriptstyle V}\,\alpha}^{(\lambda_\rho)}}^*\!(k^\prime)
{k^\prime}_\beta}{M_\pi^2-k^2-i0}\,
\overline{u}(p^\prime,s^\prime)\not\! k \gamma^5u(p,s),
\label{12}
\end{multline} 
where $g_{\rho\pi\gamma}$ and $g_{\pi{\scriptscriptstyle NN}}$ are 
the coupling constants related to the $\rho\pi\gamma$ and $\pi NN$ vertices. 
The vertex $\rho\pi\gamma$ is generated by an effective Lagrangian with a
minimal number of derivatives: 
\begin{equation}
{\cal L}_{\rho\pi\gamma}(x)=\frac{eg_{\rho\pi\gamma}}{4M_{\rho}}
\varepsilon^{\mu\nu\alpha\beta}F_{\mu\nu}(x)\vec \rho_{\alpha\beta}(x)\cdot
\vec\pi(x)\,. 
\label{10}
\end{equation}
For the $\pi NN$ vertex we use a pseudovector coupling with 
$f_{\pi{\scriptscriptstyle NN}}=
\frac{M_\pi}{2m_{\scriptscriptstyle N}}g_{\pi{\scriptscriptstyle NN}}$: 
\begin{equation}
{\cal L}_{\pi{\scriptscriptstyle NN}}(x)=
\frac{g_{\pi{\scriptscriptstyle NN}}}{2m_{\scriptscriptstyle N}}
\,\overline{N}(x)\gamma^\mu\gamma^5\vec\tau 
N(x)\cdot \partial_\mu\vec\pi(x)\,. 
\label{11} 
\end{equation} 
The coupling constant $g_{\rho\pi\gamma}$ is deduced from the 
$\rho^0\!\to\!\gamma+\pi^0$ decay width 
\begin{equation}
\Gamma_{\rho\!\to\!\gamma\pi}=\frac{\alpha}{3}\,
\frac{g_{\rho\pi\gamma}^2}{M_{\rho}^2}
\left(\frac{M_{\rho}^2-M_\pi^2}{2M_{\rho}}\right)^3.
\label{13}
\end{equation}
With the experimental value of 
$\Gamma_{\rho\!\to\!\gamma\pi}=$ 93$\pm$19 keV~\cite{pdg} we get   
\begin{equation}
g_{\rho\pi\gamma}/M_{\rho}=0.658/M_{\rho}=0.848\, GeV^{-1}.
\label{14}
\end{equation}
For the $\pi NN$ coupling constant
we use the standard value of $g_{\pi{\scriptscriptstyle NN}}=$ 13.4. 

For the $\eta$ and $\eta^\prime$ mesons we have correspondingly: 
\begin{eqnarray}
g_{\rho\eta\gamma}/M_{\rho}=1.230/M_{\rho}=1.585\, GeV^{-1},\nonumber\\
g_{\rho\eta^\prime\gamma}/M_{\rho}=1.052/M_{\rho}=1.356\, GeV^{-1},
\label{14a}
\end{eqnarray} 
using $\Gamma_{\rho\!\to\!\gamma\eta}=$ 45 $\pm$ 3 keV and 
$\Gamma_{\eta^\prime\!\to\!\rho\gamma}=$ 60 $\pm$ 6 keV~\cite{pdg}. 
For the strong couplings $g_{\eta NN}$ and $g_{\eta^\prime NN}$ 
the SU(3) relation is used 
\eq 
g_{\eta_{\,8} {\scriptscriptstyle NN}}=\frac{3F-D}{\sqrt{3}(F+D)}
g_{\pi {\scriptscriptstyle NN}}
\en 
for the mixing angle 
$\theta_P\approx\,-$10$^\circ$~\cite{pdg} defining the 
state $\eta^\prime =\eta_1\cos\theta_P + \eta_8\sin\theta_P$ 
(see e.g. Refs.~\cite{kirch2,titov}). 
With $F/D=$ 0.575$\pm$0.016~\cite{hatsuda} we get 
\begin{equation}
g_{\eta{\scriptscriptstyle NN}}=4.38,
\quad g_{\eta^\prime{\scriptscriptstyle NN}}=4.34,
\label{14b}
\end{equation}
where in addition we use the ratio 
$g_{\eta_1{\scriptscriptstyle NN}}:
g_{\eta_{\,8}{\scriptscriptstyle NN}}=\sqrt{2}:1$ 
which follows from a quark-model evaluation of the non-strange 
components in $\eta_1$ and $\eta_{\,8}$.

\bigskip

\subsubsection{{\it Tensor meson exchange} ($T=f_2$, $f_2^\prime$, $a_2$)}

The Lagrangians for tensor meson interaction with nucleons and vector 
particles are constructed in the framework of tensor meson dominance 
(TMD)~\cite{renner}. 
We follow Ref.~\cite{oh} and only present necessary formulas for 
understanding the final results (see details in~\cite{oh}). 

The Lagrangian for a free tensor field $f_{\mu\nu}$ is described in the 
Fierz--Pauli framework and has a complicated form when external fields
are included. However, the equations of motion are reduced to the usual  
Klein--Gordon equations for each independent component of the symmetric tensor 
field $f_{\alpha\beta}=f_{\beta\alpha}$: 
\begin{equation}
(\partial_\mu^2-M^2_{\scriptscriptstyle T})f_{\alpha\beta}=0
\label{15}
\end{equation}
with the additional constraints  
\begin{equation}
\partial^{\,\alpha}\! f_{\alpha\beta}=0,\quad 
g^{\alpha\beta} f_{\alpha\beta}=0 \,.
\label{16}
\end{equation} 
As result there are only 5 independent components of the tensor field 
with spin 2. The propagator of the tensor field has the form 
\begin{eqnarray} 
G^{\alpha\beta;\alpha^\prime\beta^\prime}(k^2)&=&
 iP^{\alpha\beta;\alpha^\prime\beta^\prime}(k) \, 
\frac{1}{k^2-M_{\scriptscriptstyle T}^2},\\
\label{17}
P^{\alpha\beta;\alpha^\prime\beta^\prime}(k)&=&
\frac{1}{2}\left({\overline g}^{\alpha\alpha^\prime}
{\overline g}^{\beta\beta^\prime}+
{\overline g}^{\beta\alpha^\prime}{\overline g}^{\alpha\beta^\prime}\right)  
-\frac{1}{3}{\overline g}^{\alpha\beta}
{\overline g}^{\alpha^\prime\beta^\prime},\nonumber\\
{\overline g}^{\alpha\beta}&=&-g^{\alpha\beta}+
\frac{k^\alpha k^\beta}{M_{\scriptscriptstyle T}^2}\,.\nonumber
\end{eqnarray}
The electromagnetic vertex $\gamma^*+f_2\to\rho^0$ depends on 
4 Lorentz indices of the tensor and vector fields
$f_{\alpha\beta},\,A_\mu,\,\rho_\nu$ and is described as a sum of two 
independent Lorentz-covariant terms with corresponding coupling constants 
$f_{\rho f_2\gamma}$ and $g_{\rho f_2\gamma}$~\cite{renner,oh}:
\eq 
& &\Gamma^{\alpha\beta;\mu\nu}(k^\prime,q)=
\frac{ef_{\rho f_2\gamma}}{M_{f_2}^2}
\biggl[-g^{\mu\nu}q\cdot k^\prime+{k^\prime}^\nu q^\mu\biggr]\nonumber\\
&\times&(q+k^\prime)^\alpha(q+k^\prime)^\beta
+eg_{\rho f_2\gamma}\biggl[g^{\mu\nu}(q+k^\prime)^\alpha(q+k^\prime)^\beta 
\nonumber\\
&-&g^{\mu\alpha}{k^\prime}^\nu(q+k^\prime)^\beta
-g^{\mu\beta}{k^\prime}^\nu(q+k^\prime)^\alpha
-g^{\nu\alpha}q^\mu(q+k^\prime)^\beta\nonumber\\
&-&g^{\nu\beta}
q^\mu(q+k^\prime)^\alpha 
+2q\cdot k^\prime(g^{\nu\alpha}g^{\mu\beta}
+g^{\nu\beta}g^{\mu\alpha})\biggr]\,.
\label{18}
\en 
The strong coupling $f_2NN$ also includes two independent Lorentz covariant 
terms: 
\eq 
\Gamma^{\alpha\beta}(p,p^\prime)&=&\frac{G_{f_2NN}}{m_{\scriptscriptstyle N}}
\left[(p+p^\prime)^\alpha\gamma^\beta+(p+p^\prime)^\beta\gamma^\alpha\right]
\nonumber\\
&+&\frac{F_{f_2NN}}{m_{\scriptscriptstyle N}^2}(p+p^\prime)^\alpha
(p+p^\prime)^\beta \,.
\label{19}
\en 
The magnitudes of the coupling constants 
$f_{\rho f_2\gamma}$, $g_{\rho f_2\gamma}$, $F_{f_2NN}$ and 
$G_{f_2NN}$ can be estimated in the framework of VMD and TMD models 
(see details in ~\cite{renner,oh}). In particular, the couplings are expressed
in terms of two universal constants $g_{\scriptscriptstyle V}$ and 
$g_{\scriptscriptstyle T}$: 
\begin{eqnarray}
eg_{\rho f_2\gamma}&=&e\frac{g_{f_2{\scriptscriptstyle VV}}}{f_\rho M_{f_2}} 
= e \frac{g_{\scriptscriptstyle T}}{g_{\scriptscriptstyle V}M_{f_2}},\quad
G_{f_2{\scriptscriptstyle NN}}=\frac{m_{\scriptscriptstyle N}}{2M_{f_2}}
g_{\scriptscriptstyle T},\nonumber\\
f_{\rho f_2\gamma}&=&F_{f_2{\scriptscriptstyle NN}}=0,
\label{20}
\end{eqnarray}
where $f_\rho=g_{\rho\pi\pi}=g_{\scriptscriptstyle V}$, 
$g_{f_2{\scriptscriptstyle VV}}=g_{f_2\pi\pi}=g_{\scriptscriptstyle T}$. 
The diagrams in Figs.~\ref{fg3} and \ref{fg4} illustrate these conditions. 

Tensor meson exchange contributions are described by the amplitude  
\eq 
T_{f_2}(s,s^\prime,\lambda,\lambda_\rho)&=&
\frac{2eg_{\rho f_2\gamma}G_{f_2NN}}{m_{\scriptscriptstyle N}(t-M_{f_2}^2)}\,
{\cal F}_{\rho f_2\gamma}(q^2,t){\cal F}_{f_2NN}(t)\nonumber\\
&\times&
\overline{u}(p^\prime,s^\prime) \Gamma_{f_2} u(p,s) \,, 
\label{21}
\en 
where 
\eq 
\Gamma_{f_2} &=& ({\epsilon_{{\scriptscriptstyle V}}^{(\lambda_\rho)}}^*
\epsilon^{(\lambda)})\biggl[\left((p\!+\!p^\prime)(q\!+\!k^\prime)\right)
(\not \!q\!+\!\not \!k^\prime)\\
&+&\frac{2m_{\scriptscriptstyle N}}{3}
\biggl((q\!+\!k^\prime)^2-\frac{1}{M_{f_2}^2}
(k\cdot(q\!+\!k^\prime))^2\biggr)\biggr]\nonumber\\
&+&2qk^\prime\biggl[((p\!+\!p^\prime)
{\epsilon_{{\scriptscriptstyle V}}^{(\lambda_\rho)}}^* ) 
\not \!\epsilon^{(\lambda)}
+((p\!+\!p^\prime)\epsilon^{(\lambda)}) 
\not {\epsilon_{{\scriptscriptstyle V}}^{(\lambda_\rho)}}^*\nonumber\\
&+&\frac{4m_{\scriptscriptstyle N}}{3}\biggl(
({\epsilon_{{\scriptscriptstyle V}}^{(\lambda_\rho)}}^*
\epsilon^{(\lambda)})-\frac{1}{M_{f_2}^2}
({\epsilon_{{\scriptscriptstyle V}}^{(\lambda_\rho)}}^* k)
(\epsilon^{(\lambda)} k)\biggr)
\biggr] \,. \nonumber 
\en
Here we introduced the vertex form factors ${\cal F}_{\rho f_2\gamma}(q^2,t)$ 
and ${\cal F}_{f_2NN}(t)$. These modify the constants $g_{\rho f_2\gamma}$ 
and $G_{f_2NN}$ in analogy with Eqs.~(\ref{5})-- (\ref{6}) as:
\begin{eqnarray}
g_{\rho f_2\gamma}&\to& g_{\rho f_2\gamma}
{\cal F}_{\rho f_2\gamma}(Q^2,t)\,,\nonumber\\ 
G_{f_2NN}&\to& G_{f_2NN}{\cal F}_{f_2NN}(t) \,.
\label{22}
\end{eqnarray} 
The values of these constants (see Fig.~\ref{fg4}), 
$g_{\scriptscriptstyle V}$= 5.33 and $g_{\scriptscriptstyle T}$= 5.76, 
have been obtained using the decay data, $\Gamma_{\rho\to\pi\pi}=$ 150 MeV and 
$\Gamma_{f_2\to\pi\pi}=$ 156.5 MeV~\cite{pdg}, and the expressions:
\begin{eqnarray}
\Gamma_{\rho^0\to\pi\pi}&=&\frac{g_{\scriptscriptstyle V}^2}{4\pi}\,
\frac{M_\rho}{12}\left(1-\frac{4M_\pi^2}{M_\rho^2}\right)^{3/2},\nonumber\\
\Gamma_{f_2\to\pi\pi}&=&\frac{g_{\scriptscriptstyle T}^2}{4\pi}\,\frac{M_f}{20}
\left(1-\frac{4M_\pi^2}{M_f^2}\right)^{5/2} \,.
\label{23}
\end{eqnarray} 
Here we suppose ideal mixing between $f_2(1270)$ and $f_2^\prime(1525)$.
This means that the coupling of $f_2^\prime(1525)$ to the $\pi\pi$ channel 
is suppressed by the Okubo, Zweig and Iizuka rule as observed 
in experiment~\cite{pdg}. Therefore, in the first approximation 
one can neglect the contribution of $f_2^\prime(1525)$ in the 
electroproduction of $\rho^0$.

\subsubsection{Axial-vector meson exchange ($V_5=f_1$, $f_1^\prime$, $a_1$)}

Considering the $J^{PC}=1^{++}$ particle $V_5$ as a quark-antiquark 
$^3P_1$ bound state one can write the coupling $1^{++}\to\gamma^*(q)V(k^\prime)$
 in a form analogous to the $^3P_1\to\gamma^*\gamma^*$ coupling calculated 
in the quark model 
(see, e.g. \cite{cahn}):
\begin{eqnarray}
{\cal M}(1^{++}\to\gamma^*V)=
{\epsilon_{{\scriptscriptstyle V}\nu}^{(\lambda_\rho)}}^*\!
\Gamma^{\alpha,\mu\nu}_{{\scriptscriptstyle V}_5}\,
\epsilon_\mu^{(\lambda)}\epsilon_{{\scriptscriptstyle V}_5\alpha},\nonumber\\
\Gamma^{\alpha,\mu\nu}_{{\scriptscriptstyle V}_5}=
eg_{\rho {\scriptscriptstyle V}_5\gamma}\varepsilon^{\mu\nu\alpha\beta}
\frac{{k^\prime}^2 q_\beta-q^2 k^\prime_\beta}{M^2},\quad 
q^\mu\epsilon^{(\lambda)}_\mu(q)=0
\label{24b}
\end{eqnarray}
($V=\rho^0$ and $\epsilon_{{\scriptscriptstyle V}_5\alpha}$ is the
polarization vector of $V_5$).
Here it is implied that the radial derivative $R_{q\bar q}^\prime(0)$ of the
$q\bar q$ wave function is included to the constant 
$g_{\rho {\scriptscriptstyle V}_5\gamma}$ and the coupling can be further 
modified by a form factor.

Starting with this analogy we introduce the interaction Lagrangian 
\begin{multline}
{\cal L}_{\rho {\scriptscriptstyle V}_5\gamma}(x)=
e\frac{g_{\rho {\scriptscriptstyle V}_5\gamma}}{M_\rho^2}
\varepsilon^{\mu\nu\alpha\beta}V_{5\,\alpha}
(\partial_\beta A_\mu\square \rho_\nu\\
-\partial_\beta \rho_\nu
(g_{\mu\mu^\prime}\square -\partial_\mu\partial_{\mu^\prime})A^{\mu^\prime}) 
\label{24}
\end{multline} 
\begin{equation}
{\cal L}_{{\scriptscriptstyle V}_5{\scriptscriptstyle NN}}(x)=
g_{{\scriptscriptstyle V}_5{\scriptscriptstyle NN}}
\,\overline{N}(x)\gamma^\alpha\gamma^5 N(x)V_{5\,\alpha}(x), 
\label{25}
\end{equation} 
and write down the $t$--pole axial-vector meson contribution to the 
electroproduction amplitude modified by form factors: 
\eq 
\hspace*{-.25cm}
& &T_{V_5}(s,s^\prime,\lambda,\lambda_\rho)=
-eg_{\rho {\scriptscriptstyle V}_5\gamma}
g_{{\scriptscriptstyle V}_5{\scriptscriptstyle NN}}
{\cal F}_{\rho{\scriptscriptstyle V}_5 \gamma}(q^2,t)
{\cal F}_{{\scriptscriptstyle V}_5{\scriptscriptstyle NN}}(t)\nonumber\\
\hspace*{-.25cm}
&\times&\varepsilon^{\mu\nu\alpha\beta}\left(q_\nu
-\frac{q^2}{M_\rho^2}k_\nu^\prime\right)\epsilon^{(\lambda)}_\mu\!(q)
{\epsilon_{{\scriptscriptstyle V}\,\alpha}^{(\lambda_\rho)}}^*\!(k^\prime)
\nonumber\\
\hspace*{-.25cm}
&\times&\frac{g_{\beta\beta^\prime}-k_\beta k_{\beta^\prime}/M_f^2}{k^2-M_f^2}
\overline{u}(p^\prime,s^\prime)\gamma^{\beta^\prime}\gamma^5u(p,s) \,. 
\label{26}
\en 
 
On this basis the radiative decay $f_1\to\gamma\rho$ width is calculated as: 
\begin{equation}
\Gamma_{f_1\!\to\!\gamma\rho}=\frac{\alpha}{3}\,
\frac{g_{\rho f_1\gamma}^2}{M_{\rho}^2}\left(1+
\frac{M_\rho^2}{M_{f_1}^2}\right)
\left(\frac{M_{f_1}^2-M_\rho^2}{2M_{f_1}}\right)^3.
\label{27}
\end{equation}
Using the experimental value $\Gamma_{f_1\!\to\!\gamma\rho}=  
1.34 \pm 0.32$ MeV~\cite{pdg} we get: 
\begin{eqnarray}
g_{\rho f_1\gamma}/M_\rho=1.901/M_\rho=2.45 \ {\rm GeV}^{-1},\nonumber\\
g_{\rho f_1^\prime\gamma}/M_\rho=0.582/M_\rho=0.748 \ {\rm GeV}^{-1}\,. 
\label{28}
\end{eqnarray} 
To get $g_{\rho f_1^\prime\gamma}$ we have used quark 
counting~\cite{kalash3} for the matrix element of the charge operator 
\eq 
e_q=\sum_i\left(\frac{1}{2}\lambda^{(i)}_3\!+\!
\frac{1}{2\sqrt{3}}\lambda^{(i)}_8\right) 
\en 
in neutral meson--meson transitions
of opposite $C$-parity (in our case we deal with the transitions   
$f_1\to\rho^0$, $f_1^\prime\to\rho^0$ and $a_1\to\rho^0$). The dependence of the
matrix element on the isospin part of the meson ($f_n$ or $a_n$) wave function 
is described by the simple relations~\cite{kalash3}:
\begin{eqnarray}
\langle f_n(I\!=\!0)|e_q|a_n(I\!=\!1)\rangle=1,\nonumber\\
\langle f_n(I\!=\!1)|e_q|f_n(I\!=\!1)\rangle=\frac{1}{3},\nonumber\\
\langle a_n(I\!=\!0)|e_q|a_n(I\!=\!0)\rangle=\frac{1}{3}. 
\label{29}
\end{eqnarray} 
The final results for the $f_1(1285)$ and $f_1^\prime(1410)$ mesons 
\begin{multline}
\langle\rho^0|e_q|f_1^\prime\rangle/\langle\rho^0|e_q|f_1\rangle=
\sin\epsilon\,/\cos\epsilon\,,\\
g_{\rho f_1^\prime\gamma}=g_{\rho f_1\gamma}\tan\epsilon\qquad
\label{30}
\end{multline}
only depend on the mixing angle $\epsilon\approx$ 17$^\circ$~\cite{bolt} 
relating nonstrange and strange components in the initial meson 
(the final meson is the isovector $\rho^0=(\bar uu-\bar dd)/\sqrt{2}$) with  
\begin{eqnarray}
f_1(1285)&=&\cos\epsilon\frac{\bar uu+\bar dd}{\sqrt{2}}-
\sin\epsilon\,\bar ss,\nonumber\\
f_1(1410)&=&\sin\epsilon\frac{\bar uu+\bar dd}{\sqrt{2}}+
\cos\epsilon\,\bar ss.
\label{31}
\end{eqnarray} 
For the axial-vector isovector meson $a_1(1260)$ the corresponding coupling in
the electromagnetic transition $a_1+\gamma\to \rho^0$ can be expressed through 
the constant $g_{\rho f_1\gamma}$ also using relations (\ref{29}): 
$g_{\rho a_1\gamma}=\frac{1}{3}g_{\rho f_1\gamma}$. This is also 
fulfilled for any type of $f_n(a_n)$ meson considered here and we accept 
\begin{equation}
g_{\rho a_n\gamma}=\frac{1}{3}g_{\rho f_n\gamma},\quad n=J=0,\,1,\,2.
\label{32}
\end{equation}
In Ref.~\cite{kirch1} an estimate for the couplings of the $f_1(1285)$ and 
$f_1^\prime(1410)$ mesons to nucleons was obtained using the hypothesis 
of partial conservation of the axial-vector current, i.e. in analogy to the 
VMD model, which in this case is extended to neutral axial-vector mesons. 
According to Ref.~\cite{kirch1} $|g_{f_1{\scriptscriptstyle NN}}|=$ 1.46 
and $|g_{f_1^\prime{\scriptscriptstyle NN}}|=$ 10.5. If the neutral 
axial-vector current is only connected to the strange component in the
nucleon~\cite{kirch1,elis} then, following~(\ref{31}), it follows that 
these couplings have different signs and we use the values 
\begin{eqnarray}
g_{f_1{\scriptscriptstyle NN}}=-1.46,\quad 
g_{f_1^\prime{\scriptscriptstyle NN}}=10.5 \,.
\label{33}
\end{eqnarray}

%%%%%%%%%%%%%%%%%%%%%%%%%%%%%%%%%%%%%%%%%%%%%%%%%%%%%%%%%%%%%%%%%%%%%%%%
\subsection{Form factors}
%%%%%%%%%%%%%%%%%%%%%%%%%%%%%%%%%%%%%%%%%%%%%%%%%%%%%%%%%%%%%%%%%%%%%%%%

Finally in this section we make a few comments concerning the vertex form 
factors ${\cal F}_{\rho{\scriptscriptstyle M}\gamma}$ and 
${\cal F}_{\scriptscriptstyle MNN}$ showing up in expression for 
${\cal N}_{\scriptscriptstyle M}$ (see Table III). 
In the calculations we use a 
common monopole form factor describing the dependence on the virtuality of the
(absorbed) particle in the case that the other two are on the mass shell: 
\begin{multline}
{\cal F}_{\gamma M\rho^0}(Q^2,t\!=\!M_{\scriptscriptstyle M}^2)=
\frac{\Lambda_q^2}{\Lambda_q^2+Q^2},\\
{\cal F}_{{\scriptscriptstyle MNN}}(t)=\frac{\Lambda_t^2}{\Lambda_t^2-t},
\quad\Lambda_t=\Lambda_q=M_\rho.
\label{46}
\end{multline}
For the upper vertex in the diagrams of Fig.\ref{fg2} this form factor is the 
propagator of the virtual vector meson in the VMD. The same is also true for 
the form factors in the upper vertex of the analogous diagram of 
Fig.\ref{fg1}. In the interpretation of the form factor as the Fourier 
transform of the function $\Phi(y^2)$ (describing a nonlocal interaction in 
(\ref{6})) the expression of Eq.~(\ref{46}) takes only into account the 
characteristic scale $\sim\frac{1}{\Lambda}\approx\frac{1}{M_\rho}$ of the 
charge distribution of (any sort) in the hadron (but this is quite sufficient 
for our purposes). This procedure is also based on a similar description for 
quasi--elastic knockout of pions on the nucleon~\cite{neud2,faess} with 
similar kinematics. The corresponding magnitude of $\Lambda_t$ is correlated 
with data on $\pi^+$ electroproduction \cite{huber,horn,volmer}. 

If a vertex in the diagram contains two off--shell particles (as is the case 
for the upper vertex in the diagrams of Figs.~\ref{fg1} and \ref{fg2}), then a 
form factor should depend on both virtualities: $t-M_M^2$ and $Q^2$. In the 
case of pion exchange in the quasi--elastic knockout ($t\approx$ 0) the 
virtuality on t is negligible $M_\pi^2-t\approx$ 0 and the $t$--dependence in 
the $\gamma\pi\rho$ vertex can be neglected. However, in case of heavy meson 
exchange $M=f_0(a_0),f_1(a_1),f_2(a_2)$ we cannot neglect the dependence on the
virtuality $t-M_{\scriptscriptstyle M}^2$ for typical values of the momentum 
transfer squared $t\approx t_{min}$ in quasi--elastic knockout. Therefore we 
use for the $\gamma M\rho$ form factor a more complicated parametrization: 
\eq 
{\cal F}_{\gamma M\rho^0}(Q^2,t)&=&\frac{\Lambda_q^2}{\Lambda_q^2+Q^2}\,
\frac{\Lambda_0^2}{\Lambda_0^2+M_{\scriptscriptstyle M}^2-t}, 
\nonumber\\
M_{\scriptscriptstyle M}&=&M_{f_0(a_0)},\,M_{f_1(a_1)}.
\label{47}
\en 
Here the second factor is normalized to 1 for 
$t=M_{\scriptscriptstyle M}^2$ --- 
in correspondence with the normalization of the coupling $\gamma M\rho^0$ 
for the observable decay widths chosen in Eqs.~(\ref{7}) - (\ref{9}), 
(\ref{23}) and (\ref{27}) -  (\ref{28}). 
We use in the form factor~(\ref{47}) the same value for the cutoff 
$\Lambda^2_0=$ 1.2 GeV$^2/c^2$ as in Ref.~\cite{faess}. There we showed that
such a parametrization is successful to describe data on the
electroproduction of pions~\cite{huber,horn,volmer} in the framework of an
analogous $t$-pole mechanism with the off--shell $\gamma\rho\pi$ coupling. 
   
In the literature the $t$--dependence of the form factor~(\ref{47}) is usually 
represented in the form 
$\frac{\tilde{\Lambda}_0^2-M_M^2}{\tilde{\Lambda}_0^2-t}$ with approximately 
the same value for $\tilde{\Lambda}_0\approx$ 1.2--1.5 GeV$/c$. For a 
relatively small value of the meson mass $M_{\scriptscriptstyle M}\lesssim$ 1 
GeV in expression~(\ref{47}) both parametrizations lead to approximately the 
same results in the considered region $t\sim$ 0. For more massive 
mesons $M_{\scriptscriptstyle M}\gtrsim$ 1.3--1.5 GeV the value of 
$\tilde{\Lambda}_0$ will depend on the meson mass. To avoid the introduction of
new free parameters we use the parametrization~(\ref{47}) for all the 
$f_0(a_0)$ and $f_1(a_1)$ mesons. Only in the case of the $f_2$ meson we keep 
the standard parametrization 
(for the value of $\Lambda_{f_2}=$ 1.4 GeV$/c$), 
\begin{eqnarray}
{\cal F}_{\gamma f_2\rho^0}(Q^2,t)&=&\frac{\Lambda_q^2}{\Lambda_q^2+Q^2}\,
\frac{\Lambda_{f_2}^2-M_{f_2}^2}{\Lambda_{f_2}^2-t},\nonumber\\
{\cal F}_{f_2{\scriptscriptstyle NN}}(t)&=&
\frac{\Lambda_{f_2}^2-M_{f_2}^2}{\Lambda_{f_2}^2-t}\,,
\label{47a}
\end{eqnarray} 
which was already used by other authors (see e.g. Ref.~\cite{oh} and 
references therein). 

%%%%%%%%%%%%%%%%%%%%%%%%%%%%%%%%%%%%%%%%%%%%%%%%%%%%%%%%%%%%%%%%%%%%%%
\section{Electroproduction cross section: transverse and longitudinal parts} 
%%%%%%%%%%%%%%%%%%%%%%%%%%%%%%%%%%%%%%%%%%%%%%%%%%%%%%%%%%%%%%%%%%%%%%%%

Recent experiments of the CLAS~\cite{clashad,clasmor,clasphi} and 
$F_\pi$~\cite{huber,horn,volmer} Collaborations at JLAB on meson 
electroproduction in the quasi--elastic region allow in principle to separate 
individual meson exchange contributions. Therefore the corresponding 
electromagnetic and strong vertex form factors can be measured directly. 
In particular, in the CLAS experiments~\cite{clashad,clasmor,huber,horn,volmer}
the differential cross section
of meson electroproduction is separated
in longitudinal ($L$), transverse  ($T$) and mixed ($TT$, $LT$) parts as
\begin{eqnarray}
\frac{d^4\sigma}{dW^2dQ^2dtd\varphi_{\scriptscriptstyle M^\prime}}
&=&\Gamma\left\{\varepsilon\frac{d\sigma_{\scriptscriptstyle L}}{dt}+
\frac{d\sigma_{\scriptscriptstyle T}}{dt}
+\varepsilon\frac{d\sigma_{\scriptscriptstyle TT}}{dt}
\cos 2\varphi_{\scriptscriptstyle M^\prime}\right.\nonumber\\
&+&\left.\sqrt{2\varepsilon(1+\varepsilon)}
\frac{d\sigma_{\scriptscriptstyle LT}}{dt}
\cos\varphi_{\scriptscriptstyle M^\prime}\right\}
\label{34}
\end{eqnarray}
by varying $\varepsilon$ and $\phi_{\scriptscriptstyle M^\prime}$ (via the
Rosenbluth separation).

In Eq.~(\ref{34}) $W^2$ is the square of the invariant mass with
$W^2=s=(q+p)^2=(k^\prime+p^\prime)^2$; $p$ and $p^\prime$ are the 4-momenta
of the target and recoil nucleon respectively, $k^\prime$ is the 4-momentum of
the produced meson $M^\prime$ and $q$ is the 4-momentum of the virtual photon
$q=(q_0,\textbf{q})$ (see Fig..~\ref{fg1}) with $Q^2=-q^2$;
$t = (p^\prime-p)^2=(k^\prime-q)^2=k^2$ ($k$ being the 4-momentum of a virtual
meson $M$); $\phi_{\scriptscriptstyle M^\prime}$ is the angle between the
electron scattering plane and the plane spanned by the
$(\textbf{k}^\prime,\textbf{p}^\prime)$ momenta; the value of
$\Gamma=\frac{1}{(4 \pi )^2} \frac{W^2-m_{\scriptscriptstyle N}^2}
{Q^2E_e^2m_{\scriptscriptstyle N}^2}\frac{1}{1-\varepsilon}$
is the virtual photon flux. Here $E_e$ is the initial electron energy and
$\varepsilon = \left[ 1+ \frac{2 \vec{q}^2}{Q^2} tan^2 \frac{\theta_e}{2}
\right] ^{-1}$
characterizes the degree of longitudinal polarization of the virtual photon
($\theta_e$ is the angle between the momenta of the incident and scattered
electrons).

This separation permits to 
determine the contributions of $\pi$ and $\rho$ meson poles in the cross 
section of pion electroproduction ($M^\prime=\pi^+$)~\cite{faess}. In the 
reaction $p(e,e^\prime\rho^0)p$ the Rosenbluth separation~(\ref{34})
($M^\prime=\rho^0$) also increases the chances (in comparison to older less 
precise data~\cite{cassel}) to determine the contribution e.g. of the pion 
pole (see below). 

In this section we derive and present the formula for the individual 
contribution of each meson exchange considered to the longitudinal and 
transverse part of the cross section (also including the interference terms) 
using the previously shown amplitudes with a fixed photon polarization 
$\lambda=$ 0,$\,\pm\,$1. 

We start from the full amplitude as a sum of $t$-pole 
contributions of isoscalar ($f_n$) and isovector ($a_n$) mesons 
\begin{equation}
T(s,s^\prime,\lambda,\lambda_\rho)= 
\sum_{f_n,a_n}(T_{S}+T_{S_5}+T_{V_5}+T_{T}),
\label{35}
\end{equation} 
containing the expressions (\ref{3}), (\ref{12}), (\ref{21}) and (\ref{26}) 
obtained in the previous section. The original expression for the
$t$-pole amplitude $T_{\scriptscriptstyle M}$ corresponding to the exchange
of meson $M=S,\,S_5,\,V_5,\,T$ is written in general form as 
\eq 
T_{\scriptscriptstyle M}(s,s^\prime,\lambda,\lambda_\rho)&=&
{\epsilon^{(\lambda_\rho)}_\nu(k^\prime)}^*
\Gamma_{\scriptscriptstyle M}^{\varkappa,\mu\nu}
\epsilon^{(\lambda)}_\mu(q)\,
\,G_{\scriptscriptstyle M}^{\varkappa,\varkappa^\prime}(k)\nonumber\\
&\times&\overline{u}(p^\prime\!, s^\prime)
\Gamma_{\scriptscriptstyle M}^{\varkappa^\prime}u(p,s)\,,
\label{35a}
\en 
where $\Gamma_{\scriptscriptstyle M}^{\varkappa,\mu\nu}$ and 
$\Gamma_{\scriptscriptstyle M}^{\varkappa^\prime}$ are expressions for the 
$\rho M\gamma$ and $MNN$ vertices respectively (see Table II) and
$G_{\scriptscriptstyle M}^{\varkappa,\varkappa^\prime}(k)$ 
is the meson propagator.
Here it is understood that the index $\varkappa$ encodes the Lorentz indices 
of the exchanged meson $M$, i.e. $\varkappa=\alpha$ for $M=V_5$ 
(see Eq.~(\ref{24b})), 
$\varkappa=\alpha\beta$ for $M=T$, while the $\varkappa$ is omitted in the 
case of $M=S,S_5$.

After averaging and summing the probability $|T|^2$ over all polarizations 
(excluding the polarization $\lambda$ of the initial photon) with
\begin{equation}
\overline{|T^{(\lambda)}|^2}=\frac{1}{2}\sum_{s,s^\prime,\lambda_\rho}
T(s,s^\prime,\lambda,\lambda_\rho)T^*(s,s^\prime,\lambda,\lambda_\rho)
\label{36}
\end{equation} 
the separate components of the differential cross section in the Rosenbluth 
formula are reduced to the form: 
\begin{eqnarray}
\frac{d\sigma_{\scriptscriptstyle L}}{dt}&=& {\cal N}
\frac{1}{4\pi}\overline{\vert T^{(\lambda\!=\!0)}\vert^2}, 
\nonumber\\
\frac{d\sigma_{\scriptscriptstyle T}}{dt }&=& {\cal N}
\frac{1}{2}\!\sum_{\lambda\!=\pm 1}\!\frac{1}{4\pi} 
\overline{\vert T^{(\lambda)}\vert^2}\,,\nonumber\\
\frac{d\sigma_{\scriptscriptstyle TT}}{dt}&=&{\cal N}
\biggl\{-\,\frac{1}{2}\!\sum_{\lambda\!=\pm 1}\frac{1}{4\pi}
\overline{T^{(\lambda)}{T^{(\!-\!\lambda)}}^\ast}\biggr\}\,, 
\label{37}\\ 
\frac{d\sigma_{\scriptscriptstyle LT}}{dt}&=& {\cal N}\biggl\{-\,
\frac{1}{2}\!\sum_{\lambda\!=\pm 1}\!\lambda
\biggl(\frac{\overline{T^{(0)}{T^{(\lambda)}}^\ast}
+\overline{T^{(\lambda)}{T^{(0)}}^\ast}}{4\pi\sqrt{2}}\biggr)
\!\biggr\}\,. \nonumber 
\end{eqnarray} 
Here we introduce the standard constant 
$${\cal N}=\left[2m_{\scriptscriptstyle N} Q
\sqrt{1+\left(\frac{W^2-m_N^2+Q^2}{2m_N Q}\right)^2}
(W^2-m_{\scriptscriptstyle N}^2)\right]^{-1}$$
corresponding to the normalization 
of the cross section to unit flow of virtual photons. 

The individual meson $M=f_n,\,a_n$ contributions to the cross section 
(\ref{37}) can be presented in a general 
form --- in the form of products of the polarization vectors 
$\epsilon^{(\lambda)}_\mu{\epsilon^{(\lambda)}}^*_{\mu^\prime}$ with  
five independent tensors: 
$k^\mu k^{\mu^\prime}$, $p^\mu p^{\mu^\prime}$, 
$p^\mu k^{\mu^\prime}+k^\mu p^{\mu^\prime}$, $g^{\mu\mu^\prime}$ and 
$\varepsilon^{\mu\nu\alpha\beta}
\varepsilon^{\mu^\prime\nu^\prime\alpha^\prime\beta^\prime}
q_\nu q_{\nu^\prime}k_\alpha k_{\alpha^\prime}p_\beta p_{\beta^\prime}$
(tensors of the form $q^\mu k^{\mu^\prime}$, $q^\mu p^{\mu^\prime}$, etc. 
can be omitted because of the condition
$q^\mu\epsilon^{(\lambda)}_\mu=$ 0). In the lab frame   
with $p^\mu=\{m_{\scriptscriptstyle N},0,0,0\}$ the latter tensor, after 
contraction with 
$\epsilon^{(\lambda)}_\mu{\epsilon^{(\lambda)}_{\mu^\prime}}^*$, 
is transformed into the mixed product of 3-vectors: 
\eq 
& &\left\{\varepsilon^{\mu\nu\alpha\beta}
\varepsilon^{\mu^\prime\nu^\prime\alpha^\prime\beta^\prime}
\epsilon^{(\lambda)}_\mu{\epsilon^{(\lambda)}_{\mu^\prime}}^*
q_\nu q_{\nu^\prime}k_\alpha k_{\alpha^\prime}
p_\beta p_{\beta^\prime}\right\}_{\rm lab.}\nonumber\\
&=&\lambda^2m_N^2([{\bf q}\times{\bf k}]\cdot{\bm\epsilon}^{(\lambda)})
([{\bf q}\times{\bf k}]\cdot{{\bm\epsilon}^{(\lambda)}}^*).
\label{38}
\en 
Using the tensor decomposition we obtain the following 
expression for the individual contribution of meson $M$: 
\eq 
\hspace{-5mm}&&\overline{|T^{(\lambda)}_{\scriptscriptstyle M}|^2}\equiv
\frac{1}{2}\sum_{ss^\prime}\sum_{\lambda_\rho}
|T_{\scriptscriptstyle M}(s,s^\prime,\lambda,\lambda_\rho)|^2\nonumber\\
\hspace{-5mm}&&={\cal N}^2_{\scriptscriptstyle M}
\biggl\{A_{\scriptscriptstyle M} 
(\epsilon^{(\lambda)}{\epsilon^{(\lambda)}}^*) +
B_{\scriptscriptstyle M}\frac{1}{m_{\scriptscriptstyle N}^2} \, 
(k\epsilon^{(\lambda)})(k{\epsilon^{(\lambda)}}^*)\nonumber\\ 
\hspace{-5mm}&&+C_{\scriptscriptstyle M}\frac{1}{m_{\scriptscriptstyle N}^2} 
(p\epsilon^{(\lambda)}) (p{\epsilon^{(\lambda)}}^*)  \nonumber\\
\hspace{-5mm}&&+D_{\scriptscriptstyle M}\frac{1}{m_{\scriptscriptstyle N}^2}
[(p\epsilon^{(\lambda)})(k{\epsilon^{(\lambda)}}^*) +
  (k\epsilon^{(\lambda)})(p{\epsilon^{(\lambda)}}^*) ]
\nonumber\\
\hspace{-5mm}&&+E_{\scriptscriptstyle M}\frac{1}{m_{\scriptscriptstyle N}^4}
 [{\bf k}\!\times\!{\bf q}]\!\cdot\!{\bm\epsilon}^{(\lambda)} \, 
 [{\bf k}\!\times\!{\bf q}]\!\cdot\!{{\bm\epsilon}^{(\lambda)}}^*\biggr\}\,,
\label{39}
\en 
where the coefficients $A_{\scriptscriptstyle M}$, $B_{\scriptscriptstyle M}$, 
$C_{\scriptscriptstyle M}$, $D_{\scriptscriptstyle M}$ and 
$E_{\scriptscriptstyle M}$ are 
functions of three independent invariants $t=k^2$, $Q^2=-q_\mu^2$ and 
$s=(p+q)^2=W^2$. The full expressions are given in Table III and in the 
Appendix. 

We use dimensionless invariant variables 
\begin{eqnarray}
\xi_s&\equiv&\frac{pq}{m_{\scriptscriptstyle N}Q}=
\frac{W^2-m_{\scriptscriptstyle N}^2+Q^2}
{2m_{\scriptscriptstyle N}Q},\nonumber\\
\xi_t&\equiv&\frac{kq}{m_{\scriptscriptstyle N}Q}=
\frac{-t+M_\rho^2+Q^2}{2m_{\scriptscriptstyle N}Q},\,\,\,
\eta=\frac{-t}{4m_{\scriptscriptstyle N}^2},
\label{40}
\end{eqnarray} 
in terms of which the coefficients $A_{\scriptscriptstyle M}$, 
$B_{\scriptscriptstyle M}$, $C_{\scriptscriptstyle M}$ and 
$D_{\scriptscriptstyle M}$ 
can be expressed in the simplest form. 
Parameter $\xi_s$ has a simple physical meaning because it is proportional 
to the inverse of the Bjorken variable $x_{\scriptscriptstyle B}=
\frac{Q^2}{2pq}=\frac{Q}{2m_{\scriptscriptstyle N}\xi_s}$ (here the parameter 
$\xi_t$ has an analogous meaning in the $t$ channel 
for the virtual meson $M$). 
The factor ${\cal N}_{\scriptscriptstyle M}$, given in the last line of 
Table III, depends on the coupling constants, form factors and the meson 
propagator.

The interference terms have the same parametrization as the diagonal terms: 
\eq 
& &\overline{T^{(\lambda)}_{\scriptscriptstyle M}
{T^{(\lambda)}_{{\scriptscriptstyle M}^\prime}}^*+
T^{(\lambda)}_{{\scriptscriptstyle M}^\prime}
{T^{(\lambda)}_{\scriptscriptstyle M}}^*}\nonumber\\
&\equiv&\frac{1}{2}\sum_{ss^\prime}\sum_{\lambda_\rho}
\biggl[T_{\scriptscriptstyle M}(s,s^\prime,\lambda,\lambda_\rho)\,
{T_{{\scriptscriptstyle M}^\prime}}^*(s,s^\prime,\lambda,\lambda_\rho)
\nonumber\\
&+&T_{{\scriptscriptstyle M}^\prime}(s,s^\prime,\lambda,\lambda_\rho)\,
{T_{\scriptscriptstyle M}}^*(s,s^\prime,\lambda,\lambda_\rho)\biggr]
={\cal N}_{\scriptscriptstyle M}{\cal N}_{{\scriptscriptstyle M}^\prime}
\nonumber\\
&\times&\!\!\!\biggl\{A_{{\scriptscriptstyle MM}^\prime} \, 
(\epsilon^{(\lambda)}{\epsilon^{(\lambda)}}^*) +
B_{{\scriptscriptstyle MM}^\prime}\frac{1}{m_{\scriptscriptstyle N}^2}
 (k\epsilon^{(\lambda)})(k{\epsilon^{(\lambda)}}^*)\nonumber\\ 
&+&C_{{\scriptscriptstyle MM}^\prime} 
\frac{1}{m_N^2}(p\epsilon^{(\lambda)})( p{\epsilon^{(\lambda)}}^*)
\nonumber\\
&+&D_{{\scriptscriptstyle MM}^\prime}\frac{1}{m_N^2}
[ (p\epsilon^{(\lambda)})(k{\epsilon^{(\lambda)}}^*)+
  (k\epsilon^{(\lambda)})(p{\epsilon^{(\lambda)}}^*)]
\nonumber\\ 
&+&E_{{\scriptscriptstyle MM}^\prime}\frac{1}{m_{\scriptscriptstyle N}^4}
   [{\bf k}\!\times\!{\bf q}]\!\cdot\!{\bm\epsilon}^{(\lambda)} \, 
\, [{\bf k}\!\times\!{\bf q}]\!\cdot\!{{\bm\epsilon}^{(\lambda)}}^* \biggr\}
\label{41}
\en  
and vanish for mesons of opposite parity after averaging over the 
polarizations $\lambda_\rho,s,s^\prime$. We therefore consider only the two 
nontrivial contributions for $MM^\prime =ST$ and $MM^\prime =S_5V_5$. 
The corresponding coefficients $A,B,C,D$ and $E$ are given in Table IV. 

Such a form of the final results has to simplify the calculation of 
$\sigma_{\scriptscriptstyle L(T)}$ --- 
one only substitutes the following expressions into the r.h.s. of 
Eqs.~(\ref{39}) and (\ref{41}): 

1) For $\sigma_{\scriptscriptstyle L}$ ($\lambda=\,$0)
\begin{eqnarray}
&&\hspace{-8mm}(\epsilon^{(\lambda\!=\!0)}\epsilon^{(\lambda\!=\!0)})=1,
\nonumber\\
&&\hspace{-8mm}\frac{1}{m_{\scriptscriptstyle N}^{2}}(k\epsilon^{(\lambda\!=\!0)})(k
\epsilon^{(\lambda\!=\!0)\,\ast})
=\frac{(-2\eta+\xi_s\xi_t)^2}{1+\xi_s^2}\nonumber\\
&&\hspace{-8mm}=(\xi_t^2-4\eta)+
\frac{{\bf k}_{\rm lab}^2}{m_{\scriptscriptstyle N}^2} 
\,\sin^2\theta_k^{\rm lab},
\nonumber\\
&&\hspace{-8mm}\frac{1}{m_{\scriptscriptstyle N}^{2}}(p\epsilon^{(\lambda\!=\!0)})
(p\epsilon^{(\lambda\!=\!0)\,\ast})=(1+\xi_s^2),
\nonumber\\
&&\hspace{-8mm}\frac{1}{m_{\scriptscriptstyle N}^{2}}\left[
(p\epsilon^{(\lambda\!=\!0)})(k\epsilon^{(\lambda\!=\!0)\,\ast})
+(p\epsilon^{(\lambda\!=\!0)\,\ast})(k\epsilon^{(\lambda\!=\!0)})\right]
\nonumber\\
&&\hspace{-8mm}=2(-2\eta+\xi_s\xi_t),
\nonumber\\
&&\hspace{-8mm}\frac{1}{m_N^4}
([{\bf q}\times{\bf k}]\cdot{\bm\epsilon}^{(\lambda\!=\!0)})
([{\bf q}\times{\bf k}]\cdot{{\bm\epsilon}^{(\lambda\!=\!0)}}^*)=0\,.
\label{42}
\end{eqnarray}
2) For $\sigma_{\scriptscriptstyle T}$ ($\lambda=\pm$1)
\begin{eqnarray}
&&\hspace{-8mm}\frac{1}{2}\sum_{\lambda=\pm 1}(\epsilon^{(\lambda)}
\epsilon^{(\lambda)\,\ast})=-1,\nonumber\\
&&\hspace{-8mm}\frac{1}{2}\sum_{\lambda=\pm 1}
\frac{1}{m_{\scriptscriptstyle N}^{2}}
(k\epsilon^{(\lambda)})(k\epsilon^{(\lambda)\,\ast})
=\frac{{\bf k}_{\rm lab}^2}{2m_{\scriptscriptstyle N}^2}\, 
\sin^2\theta_k^{\rm lab},\nonumber\\
&&\hspace{-8mm}\frac{1}{2}\sum_{\lambda=\pm 1}
\frac{1}{m_{\scriptscriptstyle N}^{2}}
(p\epsilon^{(\lambda)})(p\epsilon^{(\lambda)\,\ast})=0\,,
\nonumber\\
&&\hspace{-8mm}\frac{1}{2}\sum_{\lambda=\pm 1}\!
\frac{1}{m_{\scriptscriptstyle N}^{2}}
\left[(p\epsilon^{(\lambda)})(k\epsilon^{(\lambda)\,\ast})
+(p\epsilon^{(\lambda)\,\ast})
(k\epsilon^{(\lambda)})\right]=0\,,
\nonumber\\
&&\hspace{-8mm}\frac{1}{2}\sum_{\lambda=\pm 1}
\frac{1}{m_N^4}([{\bf q}\times{\bf k}]\cdot{\bm\epsilon}^{(\lambda)})
([{\bf q}\times{\bf k}]\cdot{{\bm\epsilon}^{(\lambda)}}^*)\nonumber\\
&&\hspace{-8mm}=(1+\xi_s^2)\frac{Q^2}{m_{\scriptscriptstyle N}^2}
\frac{{\bf k}_{\rm lab}^2}{2m_{\scriptscriptstyle N}^2}\, 
\sin^2\theta_k^{\rm lab}.
\label{43}
\end{eqnarray}
Here we use the lab frame (${\bf p}=\,$0) with the $z$ axis parallel to 
the photon momentum ${\bf q}$. Then the square of the 
3-momentum ${\bf k}$ and the energy $k_0$ of the virtual meson $M$ have 
the forms: ${\bf k}^2=4m_N^2\eta(1+\eta)$ and $k_0=\frac{t}{2m_N}=-2m_N\eta$. 
The polar angle $\theta_M=\theta_k^{\rm lab}$ 
of the virtual meson 3-momentum is only used as a variable in 
Eqs.~(\ref{42}) - (\ref{43}). It is expressed by 
values of the dimensionless parameters $\xi_s$, $\xi_t$ and $\eta$ as
\begin{eqnarray}
\frac{{\bf k}^2}{m_{\scriptscriptstyle N}^2}\sin^2\theta_k^{\rm lab}&=&
\frac{4\eta(1+\eta+\xi_s^2-\xi_s\xi_t)-\xi_t^2}{1+\xi_s^2},\nonumber\\
\frac{|{\bf k}|}{m_{\scriptscriptstyle N}}\cos\theta_k^{\rm lab}&=&\,-\,
\frac{(\xi_t+2\eta\xi_s)}{\sqrt{1+\xi_s^2}}.
\label{44}
\end{eqnarray} 
The momentum ${\bf k}^\prime$ and the polar angle $\theta^\prime_{\rho}$ 
of the emitted $\rho^0$ meson can be related to the variables ${\bf k}$ and  
$\theta_k^{lab}$ using the following relations: 
\begin{eqnarray}
{\bf k^\prime}^2\sin^2\theta^{\,\prime}_\rho&=&
{\bf k}^2\sin^2\theta_k^{\rm lab},\nonumber\\
|{\bf k^\prime}|\cos\theta^\prime_\rho&=&
|{\bf q}|+|{\bf k}|\cos\theta_k^{\rm lab},
\label{45}
\end{eqnarray}
where it is understood that in the lab frame $|{\bf q}|=Q\sqrt{1+\xi_s^2}$, 
$q_0=Q\xi_s$, $k_0^\prime=Q\xi_s-2m_{\scriptscriptstyle N}\eta$ and 
${\bf k^\prime}^2={k_0^\prime}^2-M_\rho^2$.

%%%%%%%%%%%%%%%%%%%%%%%%%%%%%%%%%%%%%%%%%%%%%%%%%%%%%%%%%%%%%%%%%%%%%%
\section{Results and discussion}
%%%%%%%%%%%%%%%%%%%%%%%%%%%%%%%%%%%%%%%%%%%%%%%%%%%%%%%%%%%%%%%%%%%%%%%%

Results for the cross sections $\sigma_{\scriptscriptstyle L(T)}$
of $\rho^0$ electroproduction in comparison with the data of the CLAS 
Collaboration~\cite{clashad} are presented in Fig.~\ref{fg5} --- 
separately for transverse (right side) $\sigma_{\scriptscriptstyle T}$
and for longitudinal (left side) $\sigma_{\scriptscriptstyle L}$
parts. For $W\gtrsim$ 2 GeV (i.e. at $x_{\scriptscriptstyle B}=$ 0.31 
and 0.38 in the CLAS kinematics) the underlying mechanism of quasi--elastic 
meson knockout (quark spin--flip in the $M1$ transitions 
$\gamma_{T}^*+\pi^0(\eta,\eta^\prime)\to\rho^0$ and change of internal orbital
 momentum in the $E1$ transitions $\gamma_{T}^*+f_n(a_n)\to\rho^0,\,n=0.1.2$) 
summed over all meson exchange contributions [see Table I and solid curves in 
Figs.\ref{fg5}-\ref{fg8}] is in agreement with the data on 
$\sigma_{\scriptscriptstyle T}(Q^2,W)$. 
However, there is no such agreement for $\sigma_{\scriptscriptstyle L}$. 
As seen from Fig.~\ref{fg5} for $\sigma_{\scriptscriptstyle T}$
the pion exchange contribution is enhanced due to the interference with the 
exchanged contributions of other pseudoscalar ($S_5=\eta,\eta^\prime$) 
and axial-vector ($V_5=f_1,f_1^\prime,a_1$) mesons (curves with short--dashed 
lines in Figs.~\ref{fg5}, \ref{fg6} and \ref{fg8}) while it is suppressed in 
$\sigma_{\scriptscriptstyle L}$. It seems that a full explanation of the large
value of $\sigma_{\scriptscriptstyle L}$ 
is based on another reaction mechanism. 

We therefore conclude that the mechanism of quasi--elastic meson knockout
from the nucleon cloud (with the conversion $M_5\to\rho^0$) is only weakly 
realized in the longitudinal cross section. 
At the same time, electroproduction through scalar $f_0$ meson exchange  
could be connected to another -- diffractive -- mechanism 
(see Fig.~\ref{fg2}$a$). It seems that in this case the couplings 
$\rho f_0\gamma$ for different $f_0$ mesons must be such that their total 
contribution to the longitudinal part $\sigma_{\scriptscriptstyle L}$ is 
equivalent to the contribution of the diffractive mechanism. However, as seen 
from the results displayed in Fig.~\ref{fg5} the total contribution of 
five $f_0$ mesons, further enhanced because of interference with the other 
mesons $f_2,a_0,a_2$ of positive parity (curves with long--dashed lines in 
Fig.~\ref{fg5}), is not enough to reproduce the data on 
$\sigma_{\scriptscriptstyle L}$. 

The mismatch of theory with data on 
$\sigma_{\scriptscriptstyle L}$ is perhaps
connected with the fact that for all five $f_0$ mesons we use a universal 
$\rho f_0\gamma$ constant $g_{\rho f_0\gamma}$ justified only 
for the radiative 
decay widths of the lightest scalars: $f_0(980)\to\rho^0+\gamma$ and 
$\rho^0\to\sigma+\gamma$. The value used here $g_{\rho f_0\gamma}=$ 0.25 
corresponds to a typical scale of electromagnetic interactions of the
$f_0$ meson interpreted as a weakly bound molecular  $K\bar K$ state 
(in case of $f_0(980)$)~\cite{lyub,kalash1} or as coupled channel state 
$\pi\bar\pi+q\bar q$ with a dominant $\pi\bar\pi$ component (in case of the 
$\sigma=f_0(600)$)~\cite{sigma}. Then there must be further scalar states with 
the $q\bar q$ component as the dominant one. In many studies 
(see, e.g.~\cite{kalash2,kalash3,lyub,torn,giac}) the heavy scalar mesons 
$f_0(1370)$, $f_0(1500)$ and $f_0(1710)$ are interpreted as either $q\bar q$
$^3P_0$ states or as a mixed state including an additional glueball 
$G_{\scriptscriptstyle J},\,J=0$ with mass $\sim$ 1.7 GeV according to lattice 
calculations~~\cite{glu}. 

In the classification of the scalar mesons we follow the
$SU(3)_{\scriptscriptstyle F} \times O(3)$ scheme of Table I.
We adopt the view that the
lowest scalar nonet [$\sigma(600)$, $f_0(980)$, $a_0(980)$, $\kappa(800)$]
is described by four-quark(antiquark) $S$-wave configurations $q^2\bar q^2$
which are strongly coupled to the open 2$\pi$, 2$K$ and $\pi K$ channels.
For the lowest-lying $^3P_0$ nonet of the $q\bar q$ system we use use the
scalar states with their masses close to the averaged mass of the
other $^3P_{J=1,2}$ nonets
[i.e. to the masses of $f_1(1285)$,  $f_1^\prime(1420)$, $f_2(1270)$,
$f^\prime_2(1525)$, ..., etc.].
Since the $^3P_0$ nonet can accomodate only two isoscalar-scalar $f_0$
configurations
only two of the three observed resonances, $f_0(1370)$, $f_0(1500)$ and $f_0(1710)$,
can be described as quarkonium states.
In this case we follow the view~\cite{giac} that these $f_0$ states result from the
mixing of two scalar-isoscalar $q\bar q$ states and an additional isosingulet
glueball configuration predicted to reside in this mass regime.
It should be noted that our final results for
the $\sigma_L$ and $d\sigma_L/dt$ cross sections are not very sensitive to
the detailed mixing scheme residing in this $f_0$ sector.
Out of the three mesons $f_0(1370)$, $f_0(1500)$ and $f_0(1710)$ we may take
any two and treat them in the meson exchange diagrams as if they were quarkonium states.
Here, for simplicity, we take the two lowest scalars $f_0(1370)$ and $f_0(1500)$.

An estimate of the radiative decays of $^3P_0$ quarkonia states done 
in Refs.~\cite{kalash1,kalash2} shows that the decay width 
$f_0\to\rho^0+\gamma$ is rather large with
$\Gamma_{f_0\to\rho\gamma}=$ 125 KeV assuming a mass of $M_{f_0}=$ 0.98 GeV.
This means that the coupling constant 
should have the value $g_{\rho f_0\gamma}=$ 1.3, 
i.e. about five times larger than the value used in the calculations. 
Starting with this alternative estimate of the coupling constant 
we recalculated the cross sections $\sigma_{\scriptscriptstyle L}$ and 
$\sigma_{\scriptscriptstyle T}$ substituting for the cases of $f_0(1370)$ and 
$f_0(1500)$ the value $g_{\rho f_0\gamma}=$ 1.3 instead of 
$g_{\rho f_0\gamma}=$ 0.25. Here we suppose that the true quarkonia states lie 
above $\sim $ 1.2 - 1.3 GeV (and thus $f_0(980)$ is not a  $^3P_0$ quarkonium
state), but the behavior of the quarkonia wave function at the origin 
$R_{\bar qq}(r\to 0)$ (which defines the value of $g_{\rho f_0\gamma}$) does not 
change significantly if the mass of the $q\bar q$ system used in calculation 
of the $f_0\to\rho_0\gamma$ branching will increase from 0.98 GeV to 
1.4 -- 1.5 GeV.

In Fig.~\ref{fg6} we present the results of this recalculation 
(the notations are the same as in Fig.~\ref{fg5}). The influence of the change
of couplings on $\sigma_{\scriptscriptstyle T}$ is negligible consistent 
with a relatively small contribution of $f_0$ exchanges to 
$\sigma_{\scriptscriptstyle T}$. At the same time the 
longitudinal cross section
$\sigma_{\scriptscriptstyle L}$ is increased considerably and now theoretical 
curves shown in Fig.~\ref{fg6} are in good agreement with the 
data~\cite{clashad} within experimental errors. 

Recall that the CLAS data at $x_{\scriptscriptstyle B}=$ 0.31 and 0.38 
correspond to invariant energies $W$ mostly above the resonance region 
($W\cong$ 2 -- 2.2 GeV). At lower energies 
(i.e. at $x_{\scriptscriptstyle B}=$ 
0.45 and 0.52 in the CLAS data) theoretical predictions failed to explain the 
data (Fig.~\ref{fg7}). It is possible that the enhancements of the cross 
sections observed in the region $W\cong$ 1.95 -- 2 GeV (this region 
corresponds to $Q^2\cong$ 2.4 - 2.6 GeV$^2/c^2$ at fixed 
$x_{\scriptscriptstyle B}=$ 0.45 in Fig.~\ref{fg7}) are consistent with some 
high-mass baryon resonances. A similar enhancement is also seen in 
$\sigma_{\scriptscriptstyle T}$ at $x_{\scriptscriptstyle B}=$ 0.38 near 
$Q^2\cong$ 1.7 -- 1.8  GeV$^2/c^2$ (i.e. near $W\approx$ 1.95 GeV), but 
unfortunately the experimental uncertainties (especially for 
$\sigma_{\scriptscriptstyle L}$) are too large in this region. 
It is interesting to note that our theoretical curves represented in 
Fig.~\ref{fg7} for all the kinematical region of the CLAS experiment
are well correlated (with only one exception for 
$\sigma_{\scriptscriptstyle L}$ at $x_{\scriptscriptstyle B}=$ 0.31) with the 
theoretical curves of Ref.~\cite{clashad} obtained on the basis of a Regge 
model~\cite{laget1,laget2,laget3,cano}.

The new published CLAS data at electron beam energy $E_e=$ 5.754 GeV with
full information on differential cross sections~\cite{clasmor} allow a more
detailed test of our results. In the region of quasi--elastic knockout
$|t-t_{min}|\lesssim$ (0.2 - 0.3) GeV$^2/c^2$ these results can be considered
as predictions and can be used in the analysis of differential cross
sections. In Fig..~\ref{fg8} we show the results for
$d\sigma_{\scriptscriptstyle T}(t,W,Q^2)/dt$ and
$d\sigma_{\scriptscriptstyle L}(t,W,Q^2)/dt$ calculated in the kinematics
above the resonance region ($W=$ 2 -- 2.4 GeV, $Q^2=$ 1.9 -- 2.2 GeV$^2/c^2$)
using enhanced values for $g_{\rho f_0\gamma}$ as done for the satisfactory
description of $\sigma_{\scriptscriptstyle L,T}$ (Figs.~\ref{fg6} -- \ref{fg7}).
The transverse cross section $d\sigma_{\scriptscriptstyle T}/dt$ largely depends
on the sign of the interference term between pseudoscalar- and
pseudovector-meson exchange contributions (the last column of Table IV), and
thus we show in Fig.~\ref{fg8} two cases: destructive (solid lines) and
constructive (dashed lines) interference of the $S_5$ and $V_5$ contributions.
As can be seen from Fig.~\ref{fg8} the variant with destructive interference
correlates well with the CLAS data on both
$d\sigma_{\scriptscriptstyle L}/dt$ and $d\sigma_{\scriptscriptstyle T}/dt$ at
small $|t|$ close to the quasi--elastic knockout region. For larger values
of $|t|\gtrsim$ 1 GeV our prediction underestimates the data,
but this deviation may not greatly change the integrated cross sections
$\sigma_{\scriptscriptstyle L/T}$. For this reason, our model predictions, originally
fitted to the old CLAS data on the integrated cross sections
$\sigma_{\scriptscriptstyle L/T}$, also succeed in a satisfactory description
of the new data on $d\sigma_{\scriptscriptstyle L/T}/dt$~\cite{clasmor}.

The full analysis of the new CLAS data~\cite{clasmor} will be presented in its own right
in a separate forthcoming paper.
The analysis of this new high-precision experimental information
in terms of the above model could clarify the role of scalar mesons in the
$\rho^0$ electroproduction and, finally, could give definite constraints on
the free parameters of the effective Lagrangians: coupling constants and form
factors.

We also should comment on the possible role of the
``non-correlated'' two-pion exchange mechanism not considered here.
The explicit contribution of the
three-pion box diagram to the $\rho^0$ photoproduction was studied in
Ref.~\cite{oh} (note that the meson exchange parameters used in Ref.~\cite{oh}
are practically the same as in the present model). The calculations
performed for values of $E_\gamma=$ 2.8, 3.28, 3.55 and 3.82 GeV shown
that the contribution of this mechanism to the differential cross section becomes
comparable to other contributions only for the very forward and backward
angles, e.g. for $|t|\lesssim$ 0.1 -- 0.2 GeV$^2/c^2$. Recall that the threshold
value of $t=t^{0}_{min}$ for the $\rho$ photoproduction is very small
($|t^{0}_{min}|\approx$ 0) when compared to the electroproduction threshold value
$t^Q_{min}$ at $Q^2\gtrsim$ 1.5 -- 2 GeV$^2/c^2$ (e.g., values of
$|t^Q_{min}|\gtrsim$ 0.2 -- 0.4 GeV$^2/c^2$
are characteristic of the CLAS kinematics as can be seen from Fig.~8).
Based on the results of Ref.~\cite{oh} we therefore think that
the non-correlated two-pion exchange does not significantly
change our results at $|t|\gtrsim$ 0.2 -- 0.4 obtained for the CLAS kinematics
with $|t^Q_{min}|\gtrsim$ 0.2 -- 0.4 GeV$^2/c^2$. But we also plan
to perform an exact evaluation of the $2\pi$ contribution to
$d\sigma_{\scriptscriptstyle L/T}/dt$ in a full analysis of the
new CLAS data.

\begin{acknowledgments}  
This work was supported by the DFG under Contract
No. FA67/31-2 and No. GRK683. 
The work is partially supported by the DFG under Contract No. 
436 RUS 113/988/01 and by the grant No. 09-02-91344 of RFBR(the Russian 
Foundation for Basic Research). This research is also part of the 
European Community-Research Infrastructure Integrating Activity 
``Study of Strongly Interacting Matter'' (HadronPhysics2, 
Grant Agreement No. 227431) and of the President grant of Russia 
``Scientific Schools''  No. 871.2008.2. 
The work is partially supported by Russian Science and Innovations 
Federal Agency under contract  No 02.740.11.0238.  
\end{acknowledgments}  

\appendix

%%%%%%%%%%%%%%%%%%%%%%%%%%%%%%%%%%%%%%%%%%%%%%%%%%%%%%%%%%%%%%%%%%%%%%
\section{Coefficients $A_T$,$B_T$,$C_T$ ,$D_T$}
%%%%%%%%%%%%%%%%%%%%%%%%%%%%%%%%%%%%%%%%%%%%%%%%%%%%%%%%%%%%%%%%%%%%%%%%
The coefficients $A,\,B,\,C,\,D$ and $E$ in Eqs.(\ref{39}) and (\ref{41}) are 
polynomials in $Q$, $t$ and $W$. In particular, the coefficients 
$A_{\scriptscriptstyle T}$,$B_{\scriptscriptstyle T}$,$C_{\scriptscriptstyle T}$ and
$D_{\scriptscriptstyle T}$
which are rather lengthy and not shown in Table III 
can be written in the form: 
\begin{eqnarray}
A_{\scriptscriptstyle T}&=&\sum_{n=-2}^{4}
\left(\frac{Q}{m_{\scriptscriptstyle N}}\right)^na_n(\eta,\xi _s,\xi _t),
\nonumber\\
B_{\scriptscriptstyle T}&=&\sum_{n=-2}^{4}
\left(\frac{Q}{m_{\scriptscriptstyle N}}\right)^nb_n(\eta,\xi _s,\xi _t),
\,\dots\,,
\label{aa1}
\end{eqnarray}
where the coefficients $a_n$, $b_n,\,\dots$ are polynomials in three 
dimensionsless variables:
\begin{eqnarray}
\eta=\frac{-t}{4m_{\scriptscriptstyle N}^2},\quad
\xi_s&=&\frac{pq}{m_{\scriptscriptstyle N}Q}=
\frac{Q}{2m_{\scriptscriptstyle N}x_{\scriptscriptstyle B}},\nonumber\\
\xi_t&=&\frac{kq}{m_{\scriptscriptstyle N}Q}=
\frac{Q}{2m_{\scriptscriptstyle N}}+\frac{M_\rho^2-t}{2m_{\scriptscriptstyle N}Q}.
\label{aa2}
\end{eqnarray}
Here we use the standard designation for Bjorken's variable 
$x_{\scriptscriptstyle B}=\frac{Q^2}{2pq}$ and introduce relative values
$\mu_\rho=\frac{m_{\scriptscriptstyle N}}{M_\rho}$ and
$\mu_f=\frac{m_{\scriptscriptstyle N}}{M_f}$ to simplify formulas.
In terms of these variables
the polynomials $a_n$, $b_n$,$c_n$ and $d_n$ for $n=$ -2, -1, $\dots$, 4 
take the form:
%%%%%%%%%%%%%%%%%%%%%%%%%%%%%%%%%%%%%%%%%%%%%%%%%%%%%%%%%%%%%%%%%%%%%
\begin{eqnarray}
a_{-2}&=&-\frac{64}{9} \eta ^2 (\eta +1) \left(4 \eta  \mu _f^2+1\right){}^2,
\nonumber\\
b_{-2}&=&\frac{64}{9}\mu _{\rho }^2\, \eta ^2 (\eta +1) 
\left(4 \eta  \mu _f^2+1\right)^{2}, 
\nonumber\\
c_{-2}&=&0,\nonumber\\
d_{-2}&=&0;
\label{dm2}
\end{eqnarray}
%%%%%%%%%%%%%%%%%%%%%%%%%%%%%%%%%%%%%%%%%%%%%%%%%%%%%%%%%%%%%%%%%%
\begin{widetext}
\begin{eqnarray}
a_{-1}&=&\frac{256}{9} \eta  \left(8 \eta ^2 (\eta +1) \mu _f^4+6 \eta  (\eta
   +1) \mu _f^2+\eta +1\right) \xi _t,\nonumber\\
b_{-1}&=&-\frac{32}{9}\xi _t \eta  \left(4 \eta  \mu _f^2+1\right) 
\left\{4\left[8 \eta(\eta +1) \mu _{\rho }^2+\eta +1\right] \mu _f^2
+8 (\eta +1) \mu _{\rho }^2-3\right\},\nonumber\\
c_{-1}&=&\frac{128}{3} \eta  \left(4 \eta  \mu _f^2+1\right) \xi _t,\nonumber\\
d_{-1}&=&-\frac{64}{3} \eta  \left(4 \eta  \mu _f^2+1\right) \xi _t;
\label{dm1}
\end{eqnarray}
%%%%%%%%%%%%%%%%%%%%%%%%%%%%%%%%%%%%%%%%%%%%%%%%%%%%%%%%%%%%%%%%%%
\begin{eqnarray}
a_{0}&=&-\frac{512}{3}\mu_f^4\xi_t^2\eta^2 (\eta +1)
-\frac{128}{9}\mu_f^2\eta\left[-12\eta\xi_s(\xi_s-\xi_t)
+\xi_t^2(6\eta+9)+8\eta(\eta+1)\right]\nonumber\\ 
&-&\frac{16}{9}\left(-24\eta\xi_s(\xi_s-\xi_t)
+\xi_t^2\left(16+\eta-9\eta^2\right)+16\eta(\eta+1)\right),\nonumber\\
b_0&=&\frac{256}{9}\mu _f^4\eta  (\eta +1) \left[4 \eta  \left(5 \xi
   _t^2+2 \eta \right) \mu _{\rho }^2+3 \xi _t^2+2 \eta \right]\nonumber\\
&+&\frac{128}{9}\mu _f^2\left\{-2 \eta ^2-\xi _t^2 \eta +6 \mu _{\rho }^2 \eta
\left[-2\eta  \xi _s^2+3 \eta  \xi _t \xi _s+(2 \eta +3) \xi _t^2+2 \eta 
   (\eta +1)\right]+\eta +2 \xi _t^2\right\}\nonumber\\
&+&\frac{4}{9}\left\{9 \eta \xi _t^2-75 \xi _t^2-24 \eta +4 \mu _{\rho }^2 
\left[-24 \eta \xi _s^2+36 \eta  \xi _t \xi _s+\left(-9 \eta ^2+4 \eta
   +16\right) \xi _t^2+16 \eta  (\eta +1)\right]\right\},\nonumber\\
c_{0}&=&-\frac{64}{3} \left(8 \eta  \left(\xi _t^2+\eta \right) \mu
   _f^2+\left(3 \eta ^2 \mu _{\rho }^2+7\right) \xi _t^2+2 \eta\right),
\nonumber\\
d_{0}&=&\frac{128}{3}\mu_f^2\eta\left[-\mu_{\rho }^2\eta\xi_t(2\xi_s-\xi_t)
+2\left(\xi_t^2+\eta\right)\right]
+\frac{32}{3}\left[7 \xi _t^2+2 \eta +\eta  \mu _{\rho }^2 \xi _t \left(-2 \xi
   _s+3 \eta  \xi _t+\xi _t\right)\right];
\label{d0}
\end{eqnarray}
%%%%%%%%%%%%%%%%%%%%%%%%%%%%%%%%%%%%%%%%%%%%%%%%%%%%%%%%%%%%%%%%%
\begin{eqnarray}
a_{1}&=&\frac{512}{9}\mu_f^4\xi_t^3\eta(\eta+1)
+\frac{128}{9}\mu_f^2\xi_t
\left[-12\eta\xi_s^2+12\eta\xi_t\xi_s-(\eta-2)\xi_t^2
+8\eta(\eta+1)\right]\nonumber\\
&+&\frac{32}{9}\xi_t\left[-6\xi_t^2-24\xi_s(\xi_s-\xi_t)
+(\eta+1)(16-9\eta)\right],\nonumber\\
b_1&=&-\frac{128}{9}\mu_f^4\xi_t\left[4\mu_{\rho }^2 
\eta(\eta +1)\left(4\xi_t^2+9\eta\right)+
(\eta +1)\left(\xi_t^2+8\eta\right)\right]\nonumber\\
&-&\frac{64}{9}\mu_f^2\left\{2\mu_{\rho }^2
\left[6\eta\xi_s\left(6\xi_t^2+\eta\right)-24\eta\xi_s^2\xi_t
+(4-8\eta)\xi_t^3+\eta(22 \eta +25)\xi_t\right]+
\xi_t\left(-6\xi_s^2+6\xi_t\xi_s-3\xi_t^2+2\eta+8\right)\right\}\nonumber\\
&-&\frac{8}{9}\mu_{\rho }^2\left\{\xi_t\left[-36\eta^2-3(3\eta+16)\xi_t^2
+52\eta+64-96\xi_s^2\right]
+24\xi_s\left(6\xi_t^2+\eta\right)\right\}
-\frac{8}{3}\xi _t \left(12\xi_s^2+3\eta-6\xi_s\xi_t-25\right),\nonumber\\
c_{1}&=&\frac{128}{3}\mu_f^2\xi_t\left(\xi_t^2+4\eta\right)
+32\mu_{\rho}^2\xi_t\eta\left(\xi_t^2+4\eta\right)
+\frac{32}{3}\xi_t\left(-12 \xi_s^2-3\xi_t^2+12\xi_s\xi_t+28\right),
\nonumber\\
d_{1}&=&\frac{128}{3}\mu_f^2\mu_{\rho }^2(2\xi_s-\xi_t)
\eta\left(2\xi_t^2+\eta\right)
-\frac{16}{3}\mu_{\rho }^2\xi_t\left[(3\eta+4)\xi_t^2+2\eta(6\eta+1)\right]
+\frac{64}{3}\mu_{\rho }^2\xi_s\left(2\xi_t^2+\eta\right)\nonumber\\
&-&\frac{8}{3}\xi_t
\left[8\left(\xi_t^2+4\eta\right)\mu_f^2-24\xi_s^2-3 \xi_t^2
+18\xi_s\xi_t+56\right];
\label{d1}
\end{eqnarray}
%%%%%%%%%%%%%%%%%%%%%%%%%%%%%%%%%%%%%%%%%%%%%%%%%%%%%%%%%%%%%%%
\begin{eqnarray}
a_{2}&=&-\frac{64}{9}\mu _f^4\xi_t^4(\eta +1)
+\frac{32}{9}\mu_f^2\xi_t^2
\left\{3\left[\xi_t^2+4\xi_s(\xi_s-\xi_t)\right]-8(\eta+1)\right\}
-4\mu_{\rho }^2\eta\xi_t^2\left[\xi_t^2+4\xi_s(\xi_s-\xi_t)\right]
\nonumber\\
&-&\frac{16}{3}(2\xi_s-\xi_t)^2\left[3\xi_s(\xi_s-\xi_t)+3\eta-4\right]
-\frac{4}{9}\left(-36 \eta ^2+28 \eta +64\right),
\nonumber\\
b_2&=&\frac{128}{9}\mu_f^4(\eta+1)\left[2\mu_{\rho }^2\left(\xi_t^4+8\eta\xi_t^2
+4\eta^2\right)+\xi_t^2+2\eta\right]\nonumber\\
&+&\frac{64}{9}\mu_f^2\left\{2\mu_{\rho}^2
\left[-3\left(2\xi_s^2-\xi_t^2\right)\left(\xi_t^2+2\eta\right)
+8\eta(\eta+1)+\xi_s \left(9\xi_t^3+
30\eta\xi_t\right)\right]
-6\xi_s^2-3\xi_t^2+4\eta+6\xi_s\xi_t+4\right\}\nonumber\\
&+&\frac{4}{3}\mu_{\rho }^2\xi_s^2\left(48\xi_s^2-144\xi_s\xi_t
+129\xi_t^2+48\eta-64\right)\nonumber\\
&+&\frac{4}{9}\mu_{\rho }^2\left\{-12\xi_s\xi_t\left(9\xi_t^2+18\eta-32\right)
+4\left[-9\eta^2+9(\eta-4)\xi_t^2+16(\eta+1)\right]\right\}
+\frac{4}{3}\left(24\xi_s^2+3\eta-12\xi_s\xi_t-25\right),\nonumber\\
c_{2}&=&-\frac{128}{3}\mu_f^2\xi_t^2
-\frac{4}{3}\mu_{\rho }^2\left[-9\xi_t^4+48 \xi_s\xi_t^3+48\left(\eta
-\xi_s^2\right)\xi_t^2+48\eta^2\right]
-\frac{32}{3}\left(-12\xi_s^2+12\xi_t\xi_s-3\xi_t^2+14\right),\nonumber\\
d_{2}&=&-\frac{64}{3}\mu_f^2\mu_{\rho }^2\xi_t
\left[-\xi_t^3+2\xi_s\left(\xi _t^2+6 \eta \right)-6\eta\xi_t\right]
+\frac{64}{3}\mu_f^2\xi_t^2
+\frac{4}{3}\left(-48\xi_s^2-6\xi_t^2+36\xi_s\xi_t+56\right)
\nonumber\\
&+&\frac{4}{3}\mu_{\rho }^2\left[48\xi_t\xi_s^3-96\xi_t^2\xi_s^2
+\xi_t\xi_s\left(51\xi_t^2+24\eta-64\right)
-6\xi_t^4+4(3\eta+8)\xi_t^2+24\eta^2\right];
\end{eqnarray}
%%%%%%%%%%%%%%%%%%%%%%%%%%%%%%%%%%%%%%%%%%%%%%%%%%%%%%%%%%%%%%%%%%
\begin{eqnarray}
a_{3}&=&8 \eta  \mu _{\rho }^2 \xi _t \left(\xi _t-2 \xi _s\right){}^2,
\nonumber\\
b_3&=&-\frac{256}{9}\mu_f^4\mu_{\rho }^2(\eta +1)\xi_t
\left(\xi_t^2+2\eta\right)
-\frac{64}{3}\mu_f^2\mu_{\rho }^2\xi_s
\left(-2\xi_s\xi _t+5\xi_t^2+4\eta\right)
+\frac{128}{9}\mu_f^2\mu_{\rho }^2\xi _t\left(3 \xi_t^2+\eta -2\right)
\nonumber\\
&-&\frac{24}{9}\mu_{\rho }^2\xi _s\left(-24 \xi _s^2+39\xi_t\xi_s
-15 \xi _t^2-12 \eta+16\right)
-\frac{8}{9}\mu_{\rho }^2\xi _t\left(9\eta-24\right),
\nonumber\\
c_{3}&=&8 \mu _{\rho }^2 \xi _t \left(-16 \xi _s^2+16 \xi _t \xi _s-3 \xi
   _t^2+4 \eta \right),\nonumber\\
d_{3}&=&\frac{32}{3}\mu_f^2\mu_{\rho}^2(2\xi_s-\xi_t)
\left(3\xi_t^2+4\eta\right)
-\frac{32}{3}\mu_{\rho}^2\xi_s\left(6\xi_s^2-15\xi_t\xi_s-9\xi_t^2
-3\eta+4\right)
-\frac{4}{3}\mu_{\rho}^2\xi_t\left(-9\xi_t^2+16\right);
\label{d3}
\end{eqnarray}
%%%%%%%%%%%%%%%%%%%%%%%%%%%%%%%%%%%%%%%%%%%%%%%%%%%%%%%%
\begin{eqnarray}
a_{4}&=&-4 \eta  \mu _{\rho }^2 \left(\xi _t-2 \xi _s\right){}^2,
\nonumber\\
b_4&=&\frac{64}{9}\mu_f^4\mu_{\rho }^2\xi_t^2(1+\eta)
+\frac{32}{3}\mu_f^2\mu_{\rho }^2\xi_t(2\xi_s-\xi_t)
+4\mu_{\rho }^2\xi_s(3\xi_s-2\xi_t),\nonumber\\
c_{4}&=&4\mu_{\rho}^2\left(16\xi_s^2-16\xi_t\xi_s+3\xi_t^2\right),
\nonumber\\
d_{4}&=&-\frac{32}{3}\mu_f^2\mu_{\rho }^2\xi_t(2\xi_s-\xi_t)
-4\mu_{\rho}^2\left(8\xi_s^2-7\xi_t\xi_s+\xi_t^2\right).
\label{d4}
\end{eqnarray}
%%%%%%%%%%%%%%%%%%%%%%%%%%%%%%%%%%%%%%%%%%%%%%%%%%%%%%%%%%%%%%%%%%
\end{widetext}

\newpage 

\begin{widetext}

%%%%%%%%%%%%%%%%%%%%%%%%%%%%%%%%%%%%%%%%%%%%%%%%%%%%%%%%%%%%%%%%%%%%%%%%%%%
\begin{table}
\begin{center}
{\bf Table I.} 
$SU(3)_F\times O(3)$ classification of neutral mesons contributing to the 
electroproduction of $\rho^0$ (the octet-singlet mixing is omitted for 
simplicity). Quark model (QM) and hadronic molecular (HM) 
states usually used for description of meson properties are also shown 
(including a possible scalar glueball $G_0$).
\end{center} 
\vspace*{.2cm}
\def\arraystretch{1.1}
\begin{tabular}{|l|c|c|c|} 
\hline 
$I^G(J^{PC})$ & QM ($^{2S+1}L_J$) or HM & 
$SU(3)$ octet states &  
$SU(3)$ singlet states \\[1pt]
\hline 
$0^+(0^{-+})$ & 
$^1\!S_0\qquad\quad$   & 
$\eta(540)$ & 
$\eta^\prime(958)$\\[1pt]
\hline 
$1^-(0^{-+})$ & 
$^1\!S_0\qquad\quad$  & 
$\pi(140)$ & \\[1pt]
\hline 
&&&\\[-10pt]
$0^+(0^{++})$ & 
$\qquad\quad K\bar K,\,2\pi$    & 
$f_0(980)$    & 
$f_0(600)\equiv\sigma$ \\
\hline 
$0^+(0^{++})$ & 
$^3\!P_0\qquad\quad$     & 
$f_0(1370)$   & 
$f_0(1500)$\\[1pt]
\hline 
$0^+(0^{++})$ & 
$\qquad\quad (G_0)$     &
              &$f_0(1710)$ \\[1pt]
\hline 
$0^+(1^{++})$ & 
$^3\!P_1\qquad\quad$ &   
$f_1(1285)$ & 
$f_1^\prime(1420)$ \\[1pt]
\hline 
$0^+(2^{++})$ & 
$^3\!P_2\qquad\quad$     &
$f_2(1270)$   & 
$f_2^\prime(1525)$ \\[1pt]
\hline 
&&&\\[-10pt]
$1^-(0^{++})$ &
$\qquad\quad K\bar K$  & 
$a_0(980)$ & \\ 
\hline 
$1^-(0^{++})$ &
$^3\!P_0\qquad\quad$   & 
$a_0(1450)$ & \\[1pt]
\hline 
$1^-(1^{++})$ &
$^3\!P_1\qquad\quad$   & 
$a_1(1260)$ & \\[1pt] 
\hline 
$1^-(2^{++})$ &
$^3\!P_2\qquad\quad$   & 
$a_2(1320)$ & \\[1pt] 
\hline 
\end{tabular}
%\end{table} 

\vspace*{0.5cm} 

%\begin{table}
\begin{center}

{\bf Table II.} Expressions for the $\rho M\gamma$ and $MNN$ vertices 

\vspace{0.5cm}

\begin{tabular}{|c|c|c|c|c|}
\hline
$M$&$S_5\,(\eta,\pi^0)$&$V_5\,(f_1,a_1)$&$S\,(f_0,a_0)$&$T\,(f_2,a_2)$\\[1pt]
\hline
&&&&\\[-11pt]
$\Gamma^{\varkappa,\mu\nu}_{\scriptscriptstyle M}$&
$\mbox{-}eg_{\rho{\scriptscriptstyle M}\gamma}
\varepsilon^{\mu\nu\alpha\beta}
\frac{q_\alpha{k^\prime}_{\!\!\beta}}{M_\rho}\!$&
-$eg_{\rho{\scriptscriptstyle M}\gamma}
\varepsilon^{\mu\nu\alpha\beta}
\frac{{k^\prime}^2q_\beta-q^2{k^\prime}_{\!\!\beta}}{M_\rho^2}$&
$eg_{\rho{\scriptscriptstyle M}\gamma}
\frac{qk^\prime}{M_\rho}\left(\!g^{\mu\nu}\!-\!
\frac{q^\mu {k^\prime}^\nu}{qk^\prime}\!\right)$&
$\!eg_{\rho{\scriptscriptstyle M}\gamma}\frac{1}{M_{\scriptscriptstyle T}}\!
\left[g^{\mu\nu}(q\!+\!k^\prime)^\alpha(q\!+\!k^\prime)^\beta\right.\!$\\
&&&&$\!\mbox{-}g^{\mu\alpha}
\!{k^\prime}^\nu\!(q\!+\!k^\prime)^\beta\!-\!
g^{\mu\beta}{k^\prime}^\nu\!(q\!+\!k^\prime)^\alpha\!$\\
&&&&
$\!\mbox{-}g^{\nu\alpha}q^\mu(q\!+\!k^\prime)^\beta\!-\!
g^{\nu\beta}q^\mu(q\!+\!k^\prime)^\alpha\!$\\
&&&&$\left.+2qk^\prime(g^{\nu\alpha}g^{\mu\beta}\!+\!
g^{\nu\beta}g^{\mu\alpha})\right]$\\[2pt]
\hline
&&&&\\[-11pt]
$\Gamma_{\scriptscriptstyle M}^\varkappa$&
$\frac{g_{\scriptscriptstyle MNN}}{2m_N}\!\!\not\!k\gamma^5$&
$g_{\scriptscriptstyle MNN}\gamma^\alpha\gamma^5$&
$g_{\scriptscriptstyle MNN}$&$\frac{g_{\scriptscriptstyle MNN}}{m_N}
[(p\!+\!p^\prime)^\alpha\gamma^\beta\!+\!
(p\!+\!p^\prime)^\beta\gamma^\alpha]$
\\[2pt]
\hline
\end{tabular}
\end{center}
\end{table}

\begin{table}
\begin{center}
{\bf Table III.} 
Coefficients $A_{\scriptscriptstyle M}$, $B_{\scriptscriptstyle M}$, 
$C_{\scriptscriptstyle M}$, $D_{\scriptscriptstyle M}$ 
and $E_{\scriptscriptstyle M}$ of Eq.~(\ref{39})

\vspace{1cm}

\hspace{-1.5cm}\begin{tabular}{|c|c|c|c|c|}
\hline
&&&&\\[-10pt]
$M$&$S_5\,(\eta,\pi^0)$&$V_5\,(f_1,a_1)$&$S\,(f_0,a_0)$&$T(f_2,a_2\!)\!$\\[1pt]
\hline
&&&&\\[-10pt]
$A_{\scriptscriptstyle M}$&
$-\eta\left(\xi_t^2\!-4\eta\right)\!$&
$2-\xi_s^2+\xi_s\xi_t-(\xi_t^2-4\eta)\frac{m_N^2}{M_\rho^2}
\left[\frac{m_N^2}{Q^2}(\!1+\eta)+\frac{2M_\rho^2}{M_{V5}^2}
\left(1+2\eta\frac{m_N^2}{M_{V5}^2}\right)
\right]$&$(1+\eta)\!\left(\!\frac{M_\rho^2}{m_N^2}
-\xi_t^2+4\eta\!\right)\!\!$&$^*)$\\[5pt]
&&$+z^2(1+\eta)\left[\frac{4m_N^2}{Q^2}\eta
-(\xi_t^2-4\eta)\frac{m_N^2}{M_\rho^2}\right]
+z\frac{2m_N}{Q}\left[2\eta\xi_s-(1+2\eta)\xi_t\right]$&&\\[4pt]
\hline
&&&&\\[-10pt]
$B_{\scriptscriptstyle M}$&$\eta$&$\frac{m_N^2}{M_\rho^2}\!
\left[\frac{m_N^2}{Q^2}(1+\eta)+\frac{2M_\rho^2}{M_{\scriptscriptstyle V5}^2}
\left(1+\!2\eta\frac{m_N^2}
{M_{\scriptscriptstyle V5}^2}\right)\right]
+z^2\!\left[\frac{m_N^2}{Q^2}(1+\!2\eta)
+\frac{m_N^2}{M_\rho^2}(1+\eta)\right]\!$&$1+\eta$&$^*)$\\[6pt]
\hline
&&&&\\[-10pt]
$C_{\scriptscriptstyle M}$&$0$&$1+z^{2}\frac{4m_N^2}{Q^2}\eta
-z\frac{2m_N}{Q}\xi_t$&$0$&$^*)$\\[4pt]
\hline
&&&&\\[-10pt]
$D_{\scriptscriptstyle M}$&$0$&$-\frac{1}{2}
-z^{2}\,\frac{2m_N^2}{Q^2}\eta+z\frac{m_{\scriptscriptstyle N}}{Q}\xi_s$&
$0$&$^*)$\\[3pt]
\hline
&&&&\\[-10pt]
$E_{\scriptscriptstyle M}$&$0$&$\frac{m_N^4}{M_\rho^2Q^2}(1+z^2)$&
$0$&$0$\\[3pt]
\hline
&&&&\\[-10pt]
$\widetilde{g}_{\rho{\scriptscriptstyle M}\gamma}\!$&
$\frac{m_{\scriptscriptstyle N}}{M_\rho}
g_{\rho{{\scriptscriptstyle S}_5}\gamma}$&
$\frac{Q^2}{zM_\rho^2}
g_{\rho{{\scriptscriptstyle V}_5}\gamma}\,,\quad z=\frac{Q^2}{M_\rho^2+Q^2}$&
$\frac{m_{\scriptscriptstyle N}}{M_\rho}
g_{\rho{\scriptscriptstyle S}\gamma}$&
$\frac{m_{\scriptscriptstyle N}}{M_{\scriptscriptstyle T}}
g_{\rho{\scriptscriptstyle T}\gamma}$\\[4pt]
\hline
${\cal N}_{\scriptscriptstyle M}$&\multicolumn{4}{c|}
{$\,\,e\widetilde{g}_{\rho{\scriptscriptstyle M}\gamma}
g_{\scriptscriptstyle MNN}
{\cal F}_{\rho{\scriptscriptstyle M}\gamma}(Q^2,t)
{\cal F}_{\scriptscriptstyle MNN}(t)\,
2m_{\scriptscriptstyle N}Q/(M_{\scriptscriptstyle M}^2-t)$}\\[3pt]
\hline
\end{tabular}
\end{center}
\vspace{-8pt}
\hspace{-12.5cm}$^*)$ see $A_{\scriptscriptstyle T}$,$B_{\scriptscriptstyle T}$,
$C_{\scriptscriptstyle T}$ and $D_{\scriptscriptstyle T}$ in Appendix
%\hspace{11cm}
\end{table}

\vspace{3cm}

\begin{table}
\begin{center}
{\bf Table IV.} $A_{{\scriptscriptstyle MM}^\prime}$, 
$B_{{\scriptscriptstyle MM}^\prime}$, $C_{{\scriptscriptstyle MM}^\prime}$, 
$D_{{\scriptscriptstyle MM}^\prime}$ and $E_{{\scriptscriptstyle MM}^\prime}$ 
of Eq.~(\ref{41})

\vspace{0.3cm}

\begin{tabular}{|c|c|c|c|}
\hline
$MM^\prime$&$ST$&$S_5V_5$\\[4pt]
\hline
$A_{{\scriptscriptstyle MM}^\prime}$&
$-8\left(\frac{Q}{m_N}\!-\!\xi_t\right)\!
\left\{\frac{Q}{m_N}(2\xi_s\!-\!\xi_t)^2
-\frac{4}{3}(1\!+\!\eta)\left[2\left(\frac{Q}{m_N}\!-\!\xi_t\right)
\right.\right.$&\\
&$\left.\left.+\eta\frac{m_N}{Q}+
\frac{m_N^2}{M_{\scriptscriptstyle T}^2}\frac{Q}{m_N}
\left(\xi_t\!-\!2\eta\frac{m_N}{Q}\right)^2
\right]\right\}$&$-(\xi_t^2\!-\!4\eta)
\left(1\!+\frac{4m_{\scriptscriptstyle N}^2}{M_{\scriptscriptstyle V5}^2}
\!\eta\right)$\\[4pt]
\hline
$B_{{\scriptscriptstyle MM}^\prime}$&
$-8\left(\frac{Q}{m_N}\!-\!\xi_t\right)
\left[\xi_s-2\frac{Q}{m_N}+\frac{4}{3}(1\!+\!\eta)
\frac{m_N^2}{M_{\scriptscriptstyle T}^2}\right]$&
$1\!+\frac{4m_{\scriptscriptstyle N}^2}{M_{\scriptscriptstyle V5}^2}
\!\eta
$\\[4pt]
\hline
$C_{{\scriptscriptstyle MM}^\prime}$&
$32\left(\frac{Q}{m_N}\!-\!\xi_t\right)^2$&$0$\\[4pt]
\hline
$D_{{\scriptscriptstyle MM}^\prime}$&
$8\left(\frac{Q}{m_N}\!-\!\xi_t\right)
\left[2\xi_s+\xi_t-2\frac{Q}{m_N}
\right]$&$0$\\[4pt]
\hline
$E_{{\scriptscriptstyle MM}^\prime}$&$0$&$0$\\[4pt]
\hline
\end{tabular}
\end{center}
\end{table}

\newpage 

%%%%%%%%%%%%%%%%%%% Fig. 1 %%%%%%%%%%%%%%%%%%%%%%%%%%%%%%%
\begin{figure}[hp]
\begin{center}
\epsfig{file=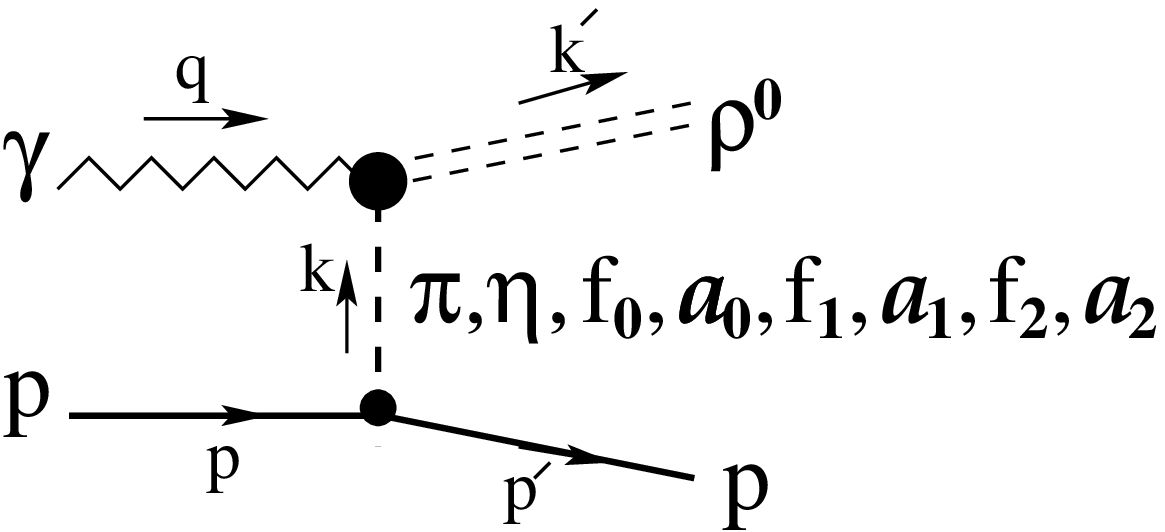,scale=.5}
\caption{$t$--pole amplitude generated by meson exchanges.} 
\label{fg1}
\end{center}
%%%%%%%%%%%%%%%%%%%%%%%%%%%%%%%%%%%%%%%%%%%%%%%%%%%%%%%%%%

%%%%%%%%%%%%%%%%%%% Fig. 2 %%%%%%%%%%%%%%%%%%%%%%%%%%%%%%%
\begin{center}
\vspace*{0.5cm}
\epsfig{file=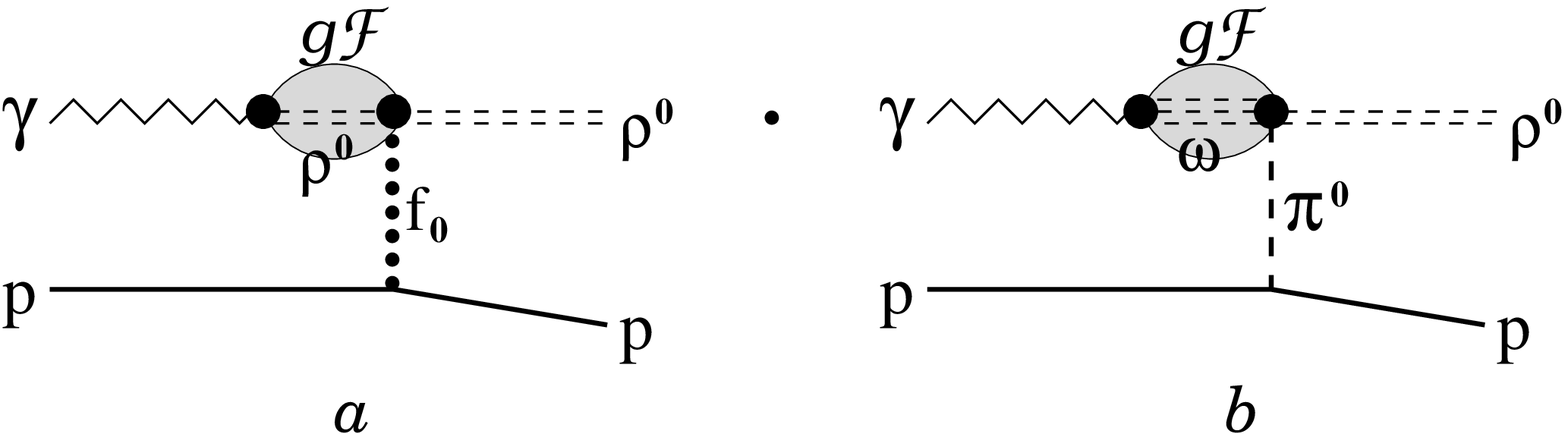,scale=.4}
\caption{Microscopic mechanism for the 
$\rho f_0 \gamma$ and $\rho\pi\gamma$ couplings.} 
\label{fg2}
\end{center}
%%%%%%%%%%%%%%%%%%%%%%%%%%%%%%%%%%%%%%%%%%%%%%%%%%%%%%%%%%

%%%%%%%%%%%%%%%%%%% Fig. 3 %%%%%%%%%%%%%%%%%%%%%%%%%%%%%%
\begin{center}
\vspace*{0.5cm}
\centering{\epsfig{figure=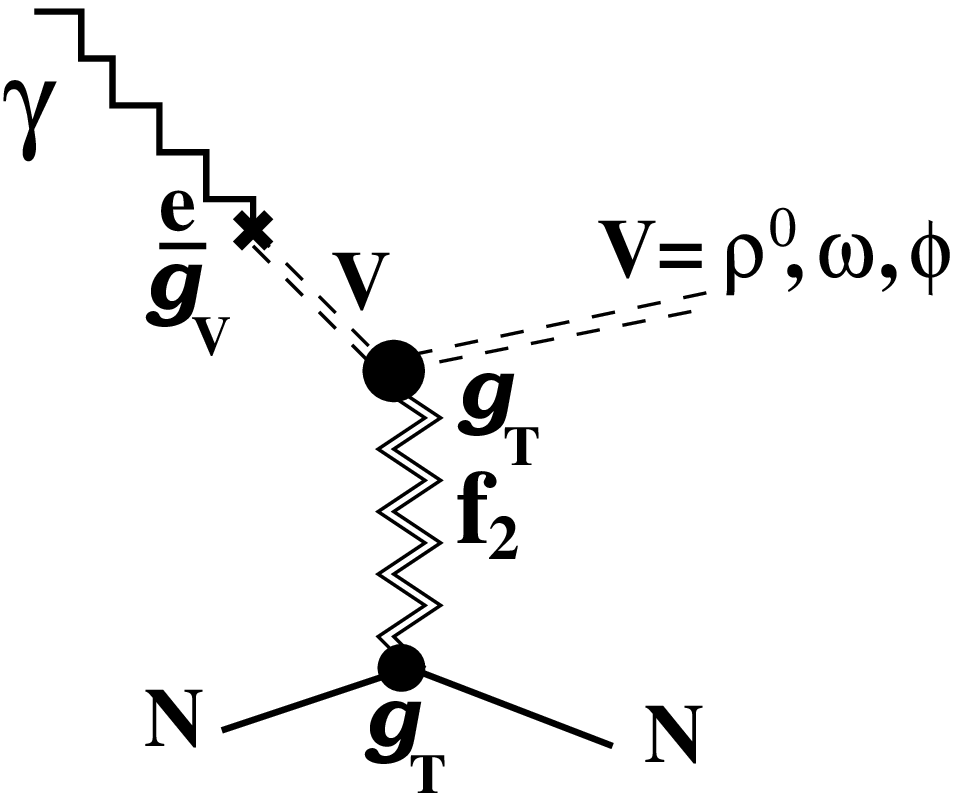,scale=0.4}}
\caption{Tensor meson contribution to the vector meson electroproduction 
in VMD and TMD models.}
\label{fg3}
\end{center}
%%%%%%%%%%%%%%%%%%%%%%%%%%%%%%%%%%%%%%%%%%%%%%%%%%%%%%%%%%%%%%%%

%%%%%%%%%%%%%%%%%%% Fig. 4 %%%%%%%%%%%%%%%%%%%%%%%%%%%%%%
\begin{center}
\vspace*{1.5cm} 
\centering{\epsfig{figure=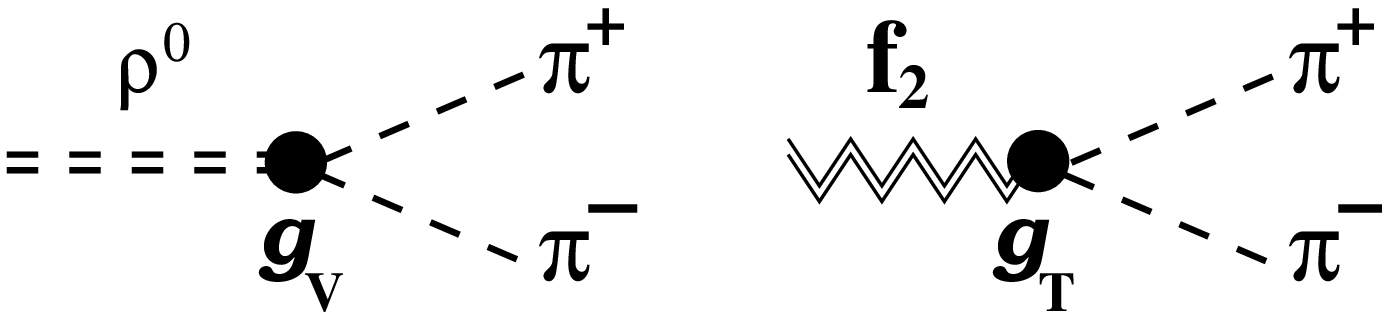,scale=0.4}}
\caption{Amplitudes of tensor meson decay into $\pi\pi$ in 
VMD and TMD models.} 
\label{fg4}
\end{center}
\end{figure}
%%%%%%%%%%%%%%%%%%%%%%%%%%%%%%%%%%%%%%%%%%%%%%%%%%%%%%%%%%%%%%%%

\newpage

%%%%%%%%%%%%%%%%%%%%% Fig. 5 %%%%%%%%%%%%%%%%%%%%%%%%%%%%%%%%%%%%%%%%
\begin{figure}[hp]
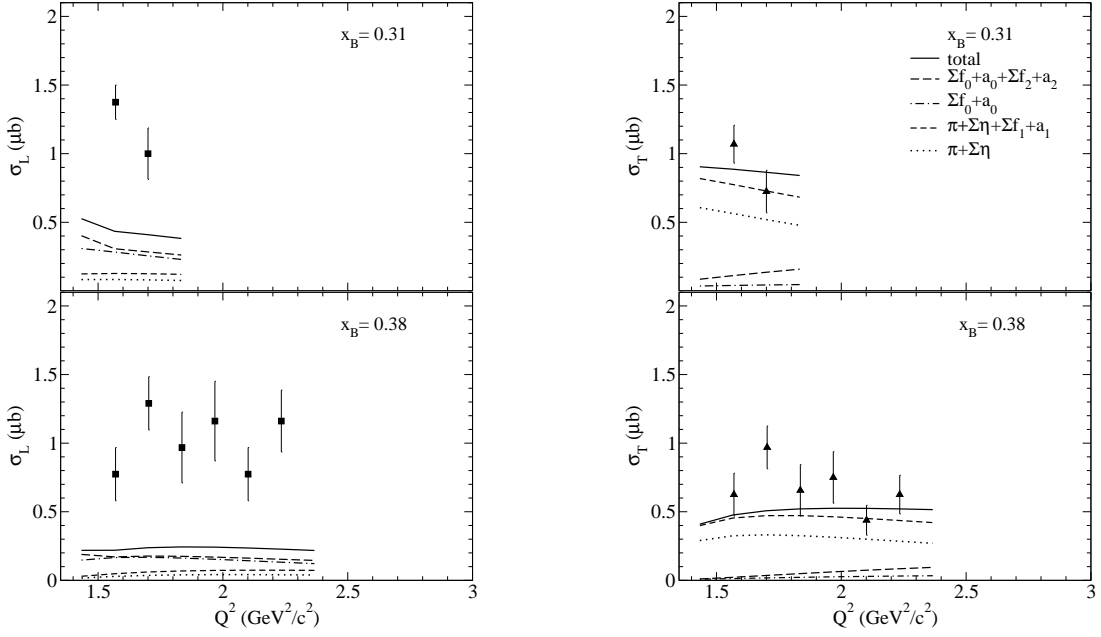


\vspace*{.5cm}

\begin{center}
\mbox{
{\epsfig{figure=fig5Lx31.eps,width=0.35\textwidth,clip}}\qquad\qquad\qquad
{\epsfig{figure=fig5Tx31.eps,width=0.35\textwidth,clip}}
}

\vspace*{-6.75mm}

\mbox{
{\epsfig{figure=fig5Lx38.eps,width=0.35\textwidth,clip}}\qquad\qquad\qquad
{\epsfig{figure=fig5Tx38.eps,width=0.35\textwidth,clip}}
}
\caption{Longitudinal (left panel) and transverse (right panel) cross
sections $\sigma_{\scriptscriptstyle L}$ and $\sigma_{\scriptscriptstyle T}$
of $\rho^0$ electroproduction as functions of $Q^2$. The sum of exchange 
contributions of all mesons listed in Table 1 is shown by solid lines.
The sum of scalar meson contributions (dashed-dotted lines) is calculated with 
a common value of ${g_{\rho f_0\gamma}}$=0.25. The sum of contributions of 
scalar and tensor mesons is shown by long-dashed lines. The sum of 
pseudoscalar and pseudovector meson contributions is shown by short-dashed 
lines (the dotted lines show the pseudoscalar meson contributions). 
Experimental values are the recent CLAS data~\cite{clashad}.}
\label{fg5}
\end{center}
\end{figure}
%%%%%%%%%%%%%%%%%%%%%%%%%%%%%%%%%%%%%%%%%%%%%%%%%%%%%%%%%%%%%%%%%%

\newpage

%%%%%%%%%%%%%%%%%%%%%%%% Fig. 6 %%%%%%%%%%%%%%%%%%%%%%%%%%%%%%%%%%
\begin{figure}[hp]
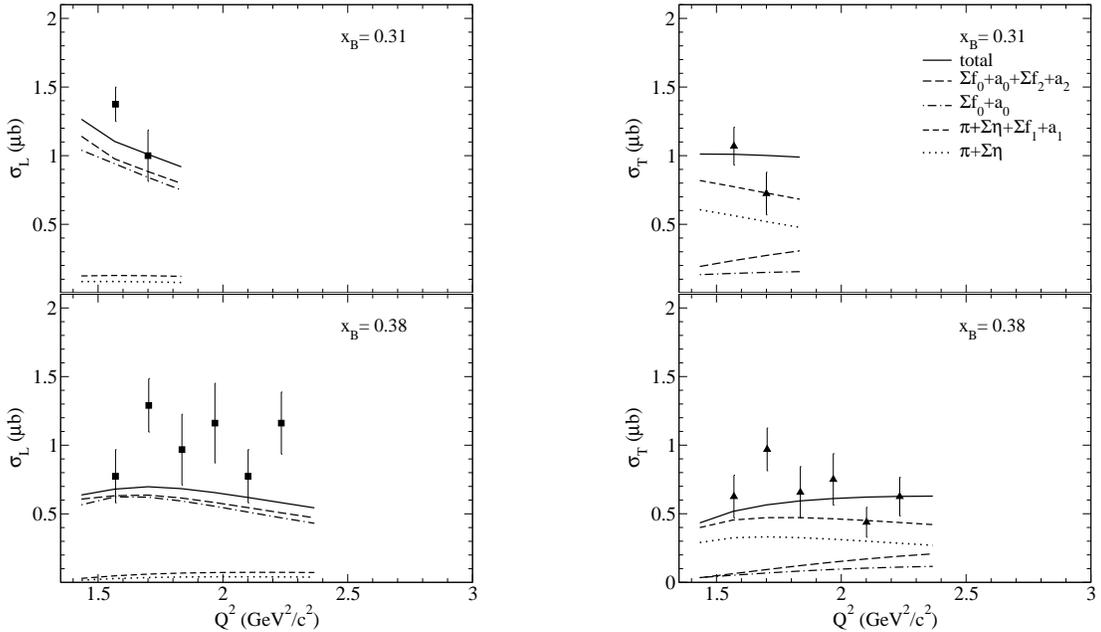

\begin{center}
\mbox{
{\epsfig{figure=fig6Lx31.eps,width=0.35\textwidth,clip}}\qquad\qquad\qquad
{\epsfig{figure=fig6Tx31.eps,width=0.35\textwidth,clip}}
}

\vspace{-6.75mm}

\mbox{
{\epsfig{figure=fig6Lx38.eps,width=0.35\textwidth,clip}}\qquad\qquad\qquad
{\epsfig{figure=fig6Tx38.eps,width=0.35\textwidth,clip}}
}
\end{center}
\caption{The same results as in Fig.~\ref{fg5} but with an enhanced 
$\rho f_0\gamma$ coupling (${g_{\rho f_0\gamma}}$= 1.3) for the two scalar 
mesons $f_0(1370)$ and $f_0(1500)$ (and for $a_0(1450)$ we use the common 
rule~(\ref{32}): $g_{\rho a_0\gamma}=\frac{1}{3}g_{\rho f_0\gamma}$).}
\label{fg6}
\end{figure}
%%%%%%%%%%%%%%%%%%%%%%%%%%%%%%%%%%%%%%%%%%%%%%%%%%%%%%%%%%

\newpage

%%%%%%%%%%%%%%%%%%%%%%%%%%% Fig. 7 %%%%%%%%%%%%%%%%%%%%%%%%%%%%%%%%%
\begin{figure}[hp]
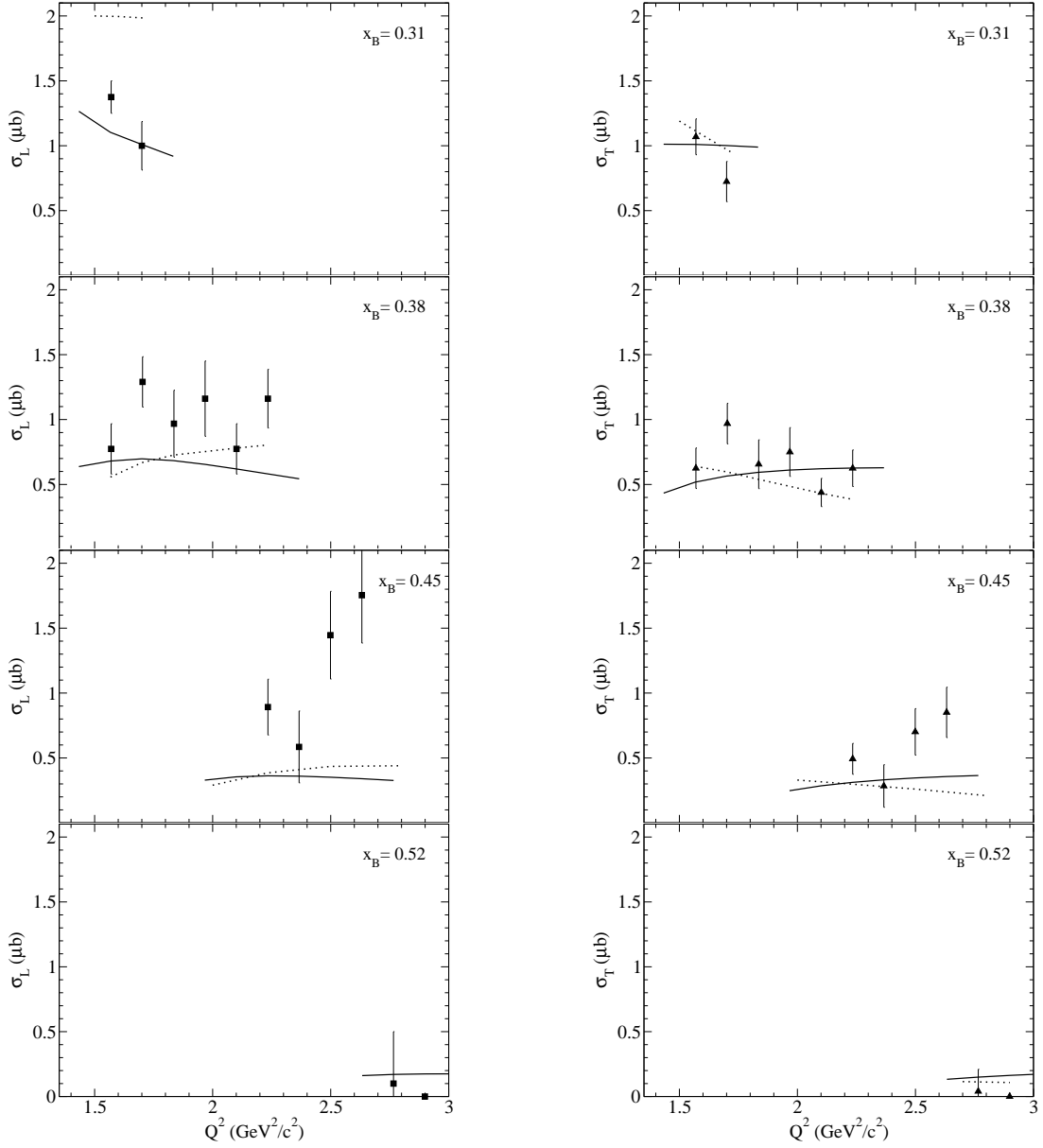

%\vspace{-10mm}
\begin{center}
\mbox{
{\epsfig{figure=fig7Lx31.eps,width=0.35\textwidth,clip}}\qquad\qquad\qquad
{\epsfig{figure=fig7Tx31.eps,width=0.35\textwidth,clip}}
}

\vspace{-6.75mm}

\mbox{
{\epsfig{figure=fig7Lx38.eps,width=0.35\textwidth,clip}}\qquad\qquad\qquad
{\epsfig{figure=fig7Tx38.eps,width=0.35\textwidth,clip}}
}

\vspace{-6.78mm}

\mbox{
{\epsfig{figure=fig7Lx45.eps,width=0.35\textwidth,clip}}\qquad\qquad\qquad
{\epsfig{figure=fig7Tx45.eps,width=0.35\textwidth,clip}}
}

\vspace{-6.75mm}

\mbox{
{\epsfig{figure=fig7Lx52.eps,width=0.35\textwidth,clip}}\qquad\qquad\qquad
{\epsfig{figure=fig7Tx52.eps,width=0.35\textwidth,clip}}
}
\end{center}
\caption{Longitudinal (left panel) and transverse (right panel) cross
sections of $\rho^0$ electroproduction. The last CLAS data~\cite{clashad} for
 $x_{\scriptscriptstyle B}=$ 0.31, 0.38, 0.45 and 0.52 are shown in comparison 
to the theoretical curves.
Solid lines correspond to the quasielastic knockout mechanism in which we
take into account the full sum of exchange diagrams for intermediate 
mesons listed in Table 1. The results~\cite{clashad} obtained  
on the basis of a Regge model of Refs.~\cite{laget1,laget2,laget3,cano}
are also shown for comparison (dotted lines).}
\label{fg7}
\end{figure}
%%%%%%%%%%%%%%%%%%%%%%%%%%%%%%%%%%%%%%%%%%%%%%%%%%%%%%%%%%%%%%%%%%%%%%%

\newpage

%%%%%%%%%%%%%%%%%%%%%%%%%%%%%%%% Fig. 8 %%%%%%%%%%%%%%%%%%%%%%%%%%%%%%%%%%%%%
\begin{figure}[hp]
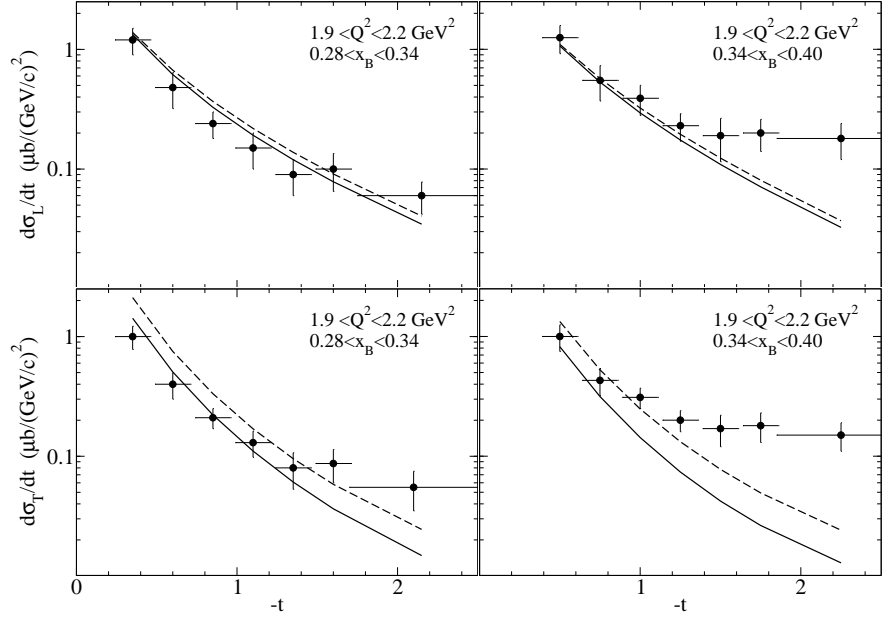

\vspace*{3cm}
\begin{center}
\mbox{
{\epsfig{figure=ds1922L2.eps,width=0.348\textwidth,clip}}\hspace{-1.2mm}
{\epsfig{figure=ds1922L3.eps,width=0.3\textwidth,clip}}
}

\vspace*{-0.6mm}

\mbox{
{\epsfig{figure=ds1922T2.eps,width=0.348\textwidth,clip}}\hspace{-1.2mm}
{\epsfig{figure=ds1922T3.eps,width=0.3\textwidth,clip}}
}
\caption{Differential cross sections $d\sigma_{\scriptscriptstyle L}/dt$ (top 
panel) and $d\sigma_{\scriptscriptstyle T}/dt$ (bottom panel) averaged over 
intervals 0.28 $<\,x_{\scriptscriptstyle B}\,<$ 0.34 (left) and 
0.34 $<\,x_{\scriptscriptstyle B}\,<$ 0.40 (right)  at
1.9 $<\,Q^2\,<$ 2.2 GeV$^2/c^2$ calculated for the same values of 
${g_{\rho f_0\gamma}}$ as used in Figs.~\ref{fg6} and \ref{fg7}. Comparison
with the latest CLAS data~\cite{clasmor} for the respective experimental bins: 
the solid and dashed lines represent the results of calculations for the cases 
of destructive and constructive interference between contributions of 
pseudoscalar ($S_5$) and pseudovector ($V_5$) mesons respectively (the 
destructive interference corresponds to the inverse sign of expressions in 
the last column of Table IV).}
\label{fg8}
\end{center}
\end{figure}
%%%%%%%%%%%%%%%%%%%%%%%%%%%%%%%%%%%%%%%%%%%%%%%%%%%%%%%%%%%%%%%%%%%%%%%%%
\end{widetext}


\begin{thebibliography}{99}
\bibitem{clashad}
C.~Hadjidakis {\it et al.} (CLAS Collaboration), 
  Phys.\ Lett.\  B {\bf 605}, 256 (2005). 
\bibitem{clasmor}
S.~A.~Morrow {\it et al.}  (CLAS Collaboration),
  Eur.\ Phys.\ J.\  A {\bf 39}, 5 (2009). 
\bibitem{clasbat}
M.~Battaglieri {\it et al.}  (CLAS Collaboration), 
  Phys.\ Rev.\ Lett.\  {\bf 87}, 172002 (2001); 
M.~Battaglieri {\it et al.}  (CLAS Collaboration),
  Phys.\ Rev.\ Lett.\  {\bf 90}, 022002 (2003). 
\bibitem{huber}
G.~M.~Huber {\it et al.}  (Jefferson Lab. $F_\pi$ Collab.), 
  Phys.\ Rev.\  C {\bf 78}, 045203 (2008). 
\bibitem{horn}T.~Horn {\it et al.}  (Jefferson Lab. $F_\pi$ Collab.), 
  Phys.\ Rev.\ Lett.\  {\bf 97}, 192001 (2006). 
\bibitem{volmer} 
J.~Volmer {\it et al.} (Jefferson Lab. $F_\pi$ Collab.), 
  Phys.\ Rev.\ Lett.\  {\bf 86}, 1713 (2001)
\bibitem{faess}
A.~Faessler, T.~Gutsche, V.~E.~Lyubovitskij and \\{} I.~T.~Obukhovsky, 
  Phys.\ Rev.\  C {\bf 76}, 025213 (2007). 
\bibitem{cassel} 
D.~G.~Cassel {\it et al.},
  %``Exclusive Rho0, Omega And Phi Electroproduction,''
  Phys.\ Rev.\  D {\bf 24}, 2787 (1981).
\bibitem{renner}
B.~Renner,
  Nucl.\ Phys.\  B {\bf 30}, 634 (1971); 
B.~Renner,
  Phys.\ Lett.\  B {\bf 33}, 599 (1970). 
\bibitem{oh}
Y.~S.~Oh and T.~S.~H.~Lee,
  Phys.\ Rev.\  C {\bf 69}, 025201 (2004). 
\bibitem{guidal}M.~Guidal and S.~Morrow, 
{\it Proceedings of the International Workshop Exclusive reactions 
at high momentum transfer, Jefferson Labaratory,
Newport-News, Virginia, USA, May 21-24 2007} (World Scientific, 2008) 
ISBN 9812796940, arXiv:0711.3743 [hep-ph].
\bibitem{kaskul} 
M.~M.~Kaskulov, K.~Gallmeister and U.~Mosel,
  Phys.\ Rev.\  D {\bf 78}, 114022 (2008);  
M.~M.~Kaskulov and U.~Mosel, 
  Phys.\ Rev.\  C {\bf 80}, 028202 (2009). 
\bibitem{laget1}
J.~M.~Laget and R.~Mendez-Galain,
  Nucl.\ Phys.\  A {\bf 581}, 397 (1995). 
\bibitem{laget2}
M.~Guidal, J.~M.~Laget and M.~Vanderhaeghen,
  Nucl.\ Phys.\  A {\bf 627}, 645 (1997). 
\bibitem{laget3}
J.~M.~Laget,
  Phys.\ Rev.\  D {\bf 70}, 054023 (2004); \\{}
J.~M.~Laget, 
  Phys.\ Lett.\  B {\bf 489}, 313 (2000); 
J.~M.~Laget,
  Nucl.\ Phys.\  A {\bf 699}, 184c (2002). 
\bibitem{cano} 
F.~Cano and J.~M.~Laget,
  Phys.\ Lett.\  B {\bf 551}, 317 (2003) 
  [Erratum-ibid.\  B {\bf 571}, 250 (2003)]. 
\bibitem{landshoff} 
A.~Donnachie and P.~V.~Landshoff, 
  Nucl.\ Phys.\  B {\bf 244}, 322 (1984); 
A.~Donnachie and P.~V.~Landshoff, arXiv:0803.0686 [hep-ph].
\bibitem{obukh}
I.~T.~Obukhovsky, D.~Fedorov, A.~Faessler, T.~Gutsche and V.~E.~Lyubovitskij, 
  Phys.\ Lett.\  B {\bf 634}, 220 (2006). 
\bibitem{neud2} 
V.~G.~Neudatchin, I.~T.~Obukhovsky, L.~L.~Sviridova and N.~P.~Yudin,
  Nucl.\ Phys.\  A {\bf 739}, 124 (2004). 
\bibitem{snd} 
 M.~N.~Achasov {\it et al.} (SND Collab.), 
  Phys.\ Lett.\  B {\bf 537}, 201 (2002). 
\bibitem{pdg} 
C.~Amsler {\it et al.}  [Particle Data Group], 
  Phys.\ Lett.\  B {\bf 667}, 1 (2008). 
\bibitem{kalash1} 
Yu.~Kalashnikova, A.~E.~Kudryavtsev, A.~V.~Nefediev, 
J.~Haidenbauer and C.~Hanhart,
  Phys.\ Rev.\  C {\bf 73}, 045203 (2006). 
\bibitem{kalash2}
 F.~E.~Close, A.~Donnachie and Yu.~S.~Kalashnikova,
  Phys.\ Rev.\  D {\bf 67}, 074031 (2003). 
\bibitem{soy} 
  B.~Friman and M.~Soyeur,
  Nucl.\ Phys.\  A {\bf 600}, 477 (1996). 
\bibitem{kissl} 
L.~S.~Kisslinger,
  %``Gluonic hadrons,''
  Nucl.\ Phys.\  A {\bf 629}, 30c (1998); \\{} 
 L.~S.~Kisslinger and W.~H.~Ma,
  Phys.\ Lett.\  B {\bf 485}, 367 (2000); 
L.~S.~Kisslinger and M.~B.~Johnson, 
  Phys.\ Lett.\  B {\bf 523}, 127 (2001). 
\bibitem{efim} 
  G.~V.~Efimov and M.~A.~Ivanov, 
  {\it The Quark Confinement Model of Hadrons}, 
  (IOP Publishing, Bristol $\&$ Philadelphia, 1993).  
\bibitem{ivanov} 
  M.~A.~Ivanov, M.~P.~Locher and V.~E.~Lyubovitskij, 
  Few Body Syst.\  {\bf 21}, 131 (1996);  
  M.~A.~Ivanov, V.~E.~Lyubovitskij, J.~G.~K\"orner and P.~Kroll, 
  Phys.\ Rev.\ D {\bf 56}, 348 (1997) 
  [arXiv:hep-ph/9612463]. 
\bibitem{fae} 
  A.~Faessler, T.~Gutsche, M.~A.~Ivanov, V.~E.~Lyubovitskij and P.~Wang,  
  Phys.\ Rev.\  D {\bf 68}, 014011 (2003). 
\bibitem{lyub} 
  T.~Branz, T.~Gutsche and V.~E.~Lyubovitskij, 
  Eur.\ Phys.\ J.\  A {\bf 37}, 303 (2008). 
\bibitem{kirch2} 
M.~Kirchbach and L.~Tiator,
  Nucl.\ Phys.\  A {\bf 604}, 385 (1996). 
\bibitem{titov} 
A.~I.~Titov, T.~S.~Lee, H.~Toki and O.~Streltsova,
  Phys.\ Rev.\  C {\bf 60}, 035205 (1999). 
\bibitem{hatsuda} 
 T.~Hatsuda,
  Nucl.\ Phys.\  B {\bf 329}, 376 (1990). 
\bibitem{cahn} 
  R.~N.~Cahn,
  Phys.\ Rev.\  D {\bf 35}, 3342 (1987). 
\bibitem{kalash3} 
F.~E.~Close, A.~Donnachie and Yu.~S.~Kalashnikova,
  %``Radiative Decays of Excited Vector Mesons,''
  Phys.\ Rev.\  D {\bf 65}, 092003 (2002). 
\bibitem{bolt} 
T.~Bolton {\it et al.}, 
  Phys.\ Lett.\  B {\bf 278}, 495 (1992).
\bibitem{kirch1} 
  M.~Kirchbach and D.~O.~Riska, 
  Nucl.\ Phys.\  A {\bf 594}, 419 (1995); 
 M.~Kirchbach, L.~Tiator, S.~Neumeier and S.~Kamalov,
  arXiv:nucl-th/9609021. 
\bibitem{elis} 
J.~R.~Ellis and M.~Karliner,
  %``Analysis of data on polarized lepton - nucleon scattering,''
  Phys.\ Lett.\  B {\bf 313}, 131 (1993). 
\bibitem{clasphi} 
 J.~P.~Santoro {\it et al.}  (CLAS Collaboration),
  %``Electroproduction of $\phi(1020)$ mesons at $1.4\leq Q^2\leq$ 3.8 GeV$^2$
  %measured with the CLAS spectrometer,''
  Phys.\ Rev.\  C {\bf 78}, 025210 (2008). 
\bibitem{sigma}
E.~van Beveren, T.~A.~Rijken, K.~Metzger, C.~Dullemond, 
G.~Rupp and J.~E.~Ribeiro,
  %``A low-lying scalar meson nonet in a unitarized meson model,''
  Z.\ Phys.\  C {\bf 30}, 615 (1986); 
E.~van Beveren and G.~Rupp,
  %``Scalar and axial-vector mesons,''
  Eur.\ Phys.\ J.\  A {\bf 31}, 468 (2007). 
\bibitem{torn}
N.~A.~Tornqvist and M.~Roos,
  %``Resurrection of the Sigma Meson,''
  Phys.\ Rev.\ Lett.\  {\bf 76}, 1575 (1996). 
\bibitem{giac}
F.~Giacosa, T.~Gutsche, V.~E.~Lyubovitskij and \\{} A.~Faessler,
  %``Scalar nonet quarkonia and the scalar glueball: mixing and decays in an
  %effective chiral approach,''
  Phys.\ Rev.\  D {\bf 72}, 094006 (2005). 
\bibitem{glu} 
Y.~Chen {\it et al.},
  %``Glueball spectrum and matrix elements on anisotropic lattices,''
  Phys.\ Rev.\  D {\bf 73}, 014516 (2006); 
J.~Sexton, A.~Vaccarino and D.~Weingarten,
  %``Coupling Constants for Scalar Glueball Decay,''
  Nucl.\ Phys.\ Proc.\ Suppl.\  {\bf 47}, 128 (1996).  
\end{thebibliography}
\end{document}